\documentclass[longauth]{aa}  

\usepackage{graphicx}
\usepackage{txfonts}
\usepackage{lipsum}
\usepackage{subcaption}         
\usepackage{lscape}             
\usepackage{placeins}           
\usepackage{orcidlink}
\usepackage{xspace}
\usepackage{natbib}
\usepackage{xcolor}
\usepackage{etoolbox}
\usepackage{bm}
\usepackage{hyperref}

\makeatletter
\pretocmd{\NAT@citex}{\color{blue}}{}{}
\makeatother

\newcommand{\XRISM}{\textit{XRISM}\xspace}
\newcommand{\XMM}{\textit{XMM-Newton}\xspace}
\newcommand{\chandra}{\textit{Chandra}\xspace}
\newcommand{\suzaku}{\textit{Suzaku}\xspace}
\newcommand{\nustar}{\textit{NuSTAR}\xspace}
\newcommand{\hitomi}{\textit{Hitomi}\xspace}

\hypersetup{
    colorlinks=true,
    linkcolor=blue,
    citecolor=blue,
    filecolor=blue,
    urlcolor=blue
}

\begin{document}

   \title{Chemical enrichment of the Perseus cluster core seen by \XRISM/Resolve}

   \author{XRISM Collaboration: Marc Audard\inst{\ref{inst1}}\orcidlink{0000-0003-4721-034X}
        \and Hisamitsu Awaki\inst{\ref{inst2}}
        \and Ralf Ballhausen\inst{\ref{inst3},\ref{inst4},\ref{inst5}}\orcidlink{0000-0002-1118-8470}
        \and Aya Bamba\inst{\ref{inst6}}\orcidlink{0000-0003-0890-4920}
        \and Ehud Behar\inst{\ref{inst7}}\orcidlink{0000-0001-9735-4873}
        \and Rozenn Boissay-Malaquin\inst{\ref{inst8},\ref{inst4},\ref{inst5}}\orcidlink{0000-0003-2704-599X}
        \and Laura Brenneman\inst{\ref{inst9}}\orcidlink{0000-0003-2663-1954}
        \and Gregory V.\ Brown\inst{\ref{inst10}}\orcidlink{0000-0001-6338-9445}
        \and Lia Corrales\inst{\ref{inst11}}\orcidlink{0000-0002-5466-3817}
        \and Elisa Costantini\inst{\ref{inst12}}\orcidlink{0000-0001-8470-749X}
        \and Renata Cumbee\inst{\ref{inst4}}\orcidlink{000-0001-9894-295X}
        \and Maria Diaz Trigo\inst{\ref{inst13}}\orcidlink{0000-0001-7796-4279}
        \and Chris Done\inst{\ref{inst14}}\orcidlink{0000-0002-1065-7239}
        \and Tadayasu Dotani\inst{\ref{inst15}}
        \and Ken Ebisawa\inst{\ref{inst15}}\orcidlink{0000-0002-5352-7178}
        \and Megan E. Eckart\inst{\ref{inst10}}\orcidlink{0000-0003-3894-5889}
        \and Dominique Eckert\inst{\ref{inst1}}\orcidlink{0000-0001-7917-3892}
        \and Satoshi Eguchi\inst{\ref{inst16}}\orcidlink{0000-0003-2814-9336}
        \and Teruaki Enoto\inst{\ref{inst17}}\orcidlink{000-0003-1244-3100}
        \and Yuichiro Ezoe\inst{\ref{inst18}}
        \and Adam Foster\inst{\ref{inst9}}\orcidlink{0000-0003-3462-8886}
        \and Ryuichi Fujimoto\inst{\ref{inst15}}\orcidlink{0000-0002-2374-7073}
        \and Yutaka Fujita\inst{\ref{inst18}}\orcidlink{0000-0003-0058-9719}
        \and Yasushi Fukazawa\inst{\ref{inst19}}\orcidlink{0000-0002-0921-8837}
        \and Kotaro Fukushima\inst{\ref{inst30}}\orcidlink{0000-0001-8055-7113}
        \and Akihiro Furuzawa\inst{\ref{inst20}}
        \and Luigi Gallo\inst{\ref{inst21}}\orcidlink{0009-0006-4968-7108}
        \and Javier A. Garc\'ia\inst{\ref{inst4},\ref{inst22}}\orcidlink{0000-0003-3828-2448}
        \and Liyi Gu\inst{\ref{inst12}}\orcidlink{0000-0001-9911-7038}
        \and Matteo Guainazzi\inst{\ref{inst23}}\orcidlink{0000-0002-1094-3147}
        \and Kouichi Hagino\inst{\ref{inst6}}\orcidlink{0000-0003-4235-5304}
        \and Kenji Hamaguchi\inst{\ref{inst8},\ref{inst4},\ref{inst5}}\orcidlink{0000-0001-7515-2779}
        \and Isamu Hatsukade\inst{\ref{inst24}}\orcidlink{0000-0003-3518-3049}
        \and Katsuhiro Hayashi\inst{\ref{inst15}}\orcidlink{0000-0001-6922-6583}
        \and Takayuki Hayashi\inst{\ref{inst8},\ref{inst4},\ref{inst5}}\orcidlink{0000-0001-6665-2499}
        \and Natalie Hell\inst{\ref{inst10}}\orcidlink{0000-0003-3057-1536}
        \and Edmund Hodges-Kluck\inst{\ref{inst4}}\orcidlink{0000-0002-2397-206X}
        \and Ann Hornschemeier\inst{\ref{inst4}}\orcidlink{0000-0001-8667-2681}
        \and Yuto Ichinohe\inst{\ref{inst25}}\orcidlink{0000-0002-6102-1441}
        \and Daiki Ishi\inst{\ref{inst15}}
        \and Manabu Ishida\inst{\ref{inst15}}
        \and Kumi Ishikawa\inst{\ref{inst18}}
        \and Yoshitaka Ishisaki\inst{\ref{inst18}}
        \and Jelle Kaastra\inst{\ref{inst12},\ref{inst26}}\orcidlink{0000-0001-5540-2822}
        \and Timothy Kallman\inst{\ref{inst4}}
        \and Erin Kara\inst{\ref{inst27}}\orcidlink{0000-0003-0172-0854}
        \and Satoru Katsuda\inst{\ref{inst28}}\orcidlink{0000-0002-1104-7205}
        \and Yoshiaki Kanemaru\inst{\ref{inst15}}\orcidlink{0000-0002-4541-1044}
        \and Richard Kelley\inst{\ref{inst4}}\orcidlink{0009-0007-2283-3336}
        \and Caroline Kilbourne\inst{\ref{inst4}}\orcidlink{0000-0001-9464-4103}
        \and Shunji Kitamoto\inst{\ref{inst29}}\orcidlink{0000-0001-8948-7983}
        \and Shogo Kobayashi\inst{\ref{inst30}}\orcidlink{0000-0001-7773-9266}
        \and Takayoshi Kohmura\inst{\ref{inst31}}\orcidlink{0000-0003-4403-4512}
        \and Aya Kubota\inst{\ref{inst32}}
        \and Maurice Leutenegger\inst{\ref{inst4}}\orcidlink{0000-0002-3331-7595}
        \and Michael Loewenstein\inst{\ref{inst3},\ref{inst4},\ref{inst5}}\orcidlink{0000-0002-1661-4029}
        \and Yoshitomo Maeda\inst{\ref{inst15}}\orcidlink{0000-0002-9099-5755}
        \and Maxim Markevitch\inst{\ref{inst4}}
        \and Hironori Matsumoto\inst{\ref{inst33}}
        \and Kyoko Matsushita\inst{\ref{inst30}}\orcidlink{0000-0003-2907-0902}
        \and Dan McCammon\inst{\ref{inst34}}\orcidlink{0000-0001-5170-4567}
        \and Brian McNamara\inst{\ref{inst35}}
        \and Fran\c{c}ois Mernier\inst{\ref{inst36},\ref{inst3},\ref{inst4},\ref{inst5}}\thanks{Corresponding author: F. Mernier, \\ e-mail: \href{mailto:francois.mernier@utoulouse.fr}{francois.mernier@utoulouse.fr}}\orcidlink{0000-0002-7031-4772}
        \and Eric D.\ Miller\inst{\ref{inst27}}\orcidlink{0000-0002-3031-2326}
        \and Jon M.\ Miller\inst{\ref{inst11}}\orcidlink{0000-0003-2869-7682}
        \and Ikuyuki Mitsuishi\inst{\ref{inst37}}\orcidlink{0000-0002-9901-233X}
        \and Misaki Mizumoto\inst{\ref{inst38}}\orcidlink{0000-0003-2161-0361}
        \and Tsunefumi Mizuno\inst{\ref{inst39}}\orcidlink{0000-0001-7263-0296}
        \and Koji Mori\inst{\ref{inst24}}\orcidlink{0000-0002-0018-0369}
        \and Koji Mukai\inst{\ref{inst8},\ref{inst4},\ref{inst5}}\orcidlink{0000-0002-8286-8094}
        \and Hiroshi Murakami\inst{\ref{inst40}}
        \and Richard Mushotzky\inst{\ref{inst3}}\orcidlink{0000-0002-7962-5446}
        \and Hiroshi Nakajima\inst{\ref{inst41}}\orcidlink{0000-0001-6988-3938}
        \and Kazuhiro Nakazawa\inst{\ref{inst37}}\orcidlink{0000-0003-2930-350X}
        \and Jan-Uwe Ness\inst{\ref{inst42}}
        \and Kumiko Nobukawa\inst{\ref{inst44}}\orcidlink{0000-0002-0726-7862}
        \and Masayoshi Nobukawa\inst{\ref{inst43}}\orcidlink{0000-0003-1130-5363}
        \and Hirofumi Noda\inst{\ref{inst45}}\orcidlink{0000-0001-6020-517X}
        \and Hirokazu Odaka\inst{\ref{inst33}}
        \and Shoji Ogawa\inst{\ref{inst15}}\orcidlink{0000-0002-5701-0811}
        \and Anna Ogorza{\l}ek\inst{\ref{inst3},\ref{inst4},\ref{inst5}}\orcidlink{0000-0003-4504-2557}
        \and Takashi Okajima\inst{\ref{inst4}}\orcidlink{0000-0002-6054-3432}
        \and Naomi Ota\inst{\ref{inst46}}\orcidlink{0000-0002-2784-3652}
        \and Stephane Paltani\inst{\ref{inst1}}\orcidlink{0000-0002-8108-9179}
        \and Robert Petre\inst{\ref{inst4}}\orcidlink{0000-0003-3850-2041}
        \and Paul Plucinsky\inst{\ref{inst9}}\orcidlink{0000-0003-1415-5823}
        \and Frederick S.\ Porter\inst{\ref{inst4}}\orcidlink{0000-0002-6374-1119}
        \and Katja Pottschmidt\inst{\ref{inst8},\ref{inst4},\ref{inst5}}\orcidlink{0000-0002-4656-6881}
        \and Kosuke Sato\inst{\ref{inst47}}\orcidlink{0000-0001-5774-1633}
        \and Toshiki Sato\inst{\ref{inst48}}
        \and Makoto Sawada\inst{\ref{inst29}}\orcidlink{0000-0003-2008-6887}
        \and Hiromi Seta\inst{\ref{inst18}}
        \and Megumi Shidatsu\inst{\ref{inst2}}\orcidlink{0000-0001-8195-6546}
        \and Aurora Simionescu\inst{\ref{inst12}}\orcidlink{0000-0002-9714-3862}
        \and Randall Smith\inst{\ref{inst9}}\orcidlink{0000-0003-4284-4167}
        \and Hiromasa Suzuki\inst{\ref{inst24}}\orcidlink{0000-0002-8152-6172}
        \and Andrew Szymkowiak\inst{\ref{inst49}}\orcidlink{0000-0002-4974-687X}
        \and Hiromitsu Takahashi\inst{\ref{inst19}}\orcidlink{0000-0001-6314-5897}
        \and Mai Takeo\inst{\ref{inst28}}
        \and Toru Tamagawa\inst{\ref{inst25}}
        \and Keisuke Tamura\inst{\ref{inst8},\ref{inst4},\ref{inst5}}
        \and Takaaki Tanaka\inst{\ref{inst50}}\orcidlink{0000-0002-4383-0368}
        \and Atsushi Tanimoto\inst{\ref{inst51}}\orcidlink{0000-0002-0114-5581}
        \and Makoto Tashiro\inst{\ref{inst28},\ref{inst15}}\orcidlink{0000-0002-5097-1257}
        \and Yukikatsu Terada\inst{\ref{inst28},\ref{inst15}}\orcidlink{0000-0002-2359-1857}
        \and Yuichi Terashima\inst{\ref{inst2}}\orcidlink{0000-0003-1780-5481}
        \and Yohko Tsuboi\inst{\ref{inst52}}
        \and Masahiro Tsujimoto\inst{\ref{inst15}}\orcidlink{0000-0002-9184-5556}
        \and Hiroshi Tsunemi\inst{\ref{inst33}}
        \and Takeshi Tsuru\inst{\ref{inst17}}\orcidlink{0000-0002-5504-4903}
        \and Ay\c{s}eg\"{u}l T\"{u}mer\inst{\ref{inst8},\ref{inst4},\ref{inst5}}\orcidlink{0000-0002-3132-8776}
        \and Hiroyuki Uchida\inst{\ref{inst17}}\orcidlink{0000-0003-1518-2188}
        \and Nagomi Uchida\inst{\ref{inst64}}\orcidlink{0000-0002-5641-745X}
        \and Yuusuke Uchida\inst{\ref{inst31}}\orcidlink{0000-0002-7962-4136}
        \and Hideki Uchiyama\inst{\ref{inst53}}\orcidlink{0000-0003-4580-4021}
        \and Shutaro Ueda\inst{\ref{inst54}}\orcidlink{0000-0001-6252-7922}
        \and Yoshihiro Ueda\inst{\ref{inst55}}\orcidlink{0000-0001-7821-6715}
        \and Shinichiro Uno\inst{\ref{inst56}}
        \and Jacco Vink\inst{\ref{inst57},\ref{inst12}}\orcidlink{0000-0002-4708-4219}
        \and Shin Watanabe\inst{\ref{inst15}}\orcidlink{0000-0003-0441-7404}
        \and Brian J.\ Williams\inst{\ref{inst4}}\orcidlink{0000-0003-2063-381X}
        \and Satoshi Yamada\inst{\ref{inst63}}\orcidlink{0000-0002-9754-3081}
        \and Shinya Yamada\inst{\ref{inst29}}\orcidlink{0000-0003-4808-893X}
        \and Hiroya Yamaguchi\inst{\ref{inst15}}\orcidlink{0000-0002-5092-6085}
        \and Kazutaka Yamaoka\inst{\ref{inst37}}\orcidlink{0000-0003-3841-0980}
        \and Noriko Yamasaki\inst{\ref{inst15}}\orcidlink{0000-0003-4885-5537}
        \and Makoto Yamauchi\inst{\ref{inst24}}\orcidlink{0000-0003-1100-1423}
        \and Shigeo Yamauchi\inst{\ref{inst58}}
        \and Tahir Yaqoob\inst{\ref{inst8},\ref{inst4},\ref{inst5}}
        \and Tomokage Yoneyama\inst{\ref{inst52}}
        \and Tessei Yoshida\inst{\ref{inst15}}
        \and Mihoko Yukita\inst{\ref{inst59},\ref{inst4}}\orcidlink{0000-0001-6366-3459}
        \and Irina Zhuravleva\inst{\ref{inst60}}\orcidlink{0000-0001-7630-8085}
        \and Elena Bellomi\inst{\ref{inst9}}\orcidlink{0000-0001-6411-3686}
        \and Ian Drury\inst{\ref{inst62}}
        \and Annie Heinrich\inst{\ref{inst60}}\orcidlink{0000-0002-7726-4202}
        \and Julie Hlavacek-Larrondo\inst{\ref{inst62}}\orcidlink{0000-0001-7271-7340}
        \and Julian Meunier\inst{\ref{inst35}}
        \and Konstantinos Migkas\inst{\ref{inst12}}\orcidlink{0000-0003-0451-0449}
        \and Lior Shefler\inst{\ref{inst61}}
        \and Phillip C. Stancil\inst{\ref{inst61}}\orcidlink{0000-0003-4661-6735}
        \and Nhut Truong\inst{\ref{inst8},\ref{inst4},\ref{inst5}}\orcidlink{0000-0003-4983-0462}
        \and Benjamin Vigneron\inst{\ref{inst62}}\orcidlink{0000-0002-2478-5119}
        \and Congyao Zhang\inst{\ref{inst64},\ref{inst60}}\orcidlink{0000-0001-5888-7052}
        \and John ZuHone\inst{\ref{inst9}}\orcidlink{0000-0003-3175-2347}
        }

   \institute{Department of Astronomy, University of Geneva, Versoix CH-1290, Switzerland\label{inst1}
            \and Department of Physics, Ehime University, Ehime 790-8577, Japan\label{inst2} 
            \and Department of Astronomy, University of Maryland, College Park, MD 20742, USA\label{inst3}
            \and NASA / Goddard Space Flight Center, Greenbelt, MD 20771, USA\label{inst4}
            \and Center for Research and Exploration in Space Science and Technology, NASA / GSFC (CRESST II), Greenbelt, MD 20771, USA\label{inst5}
            \and Department of Physics, University of Tokyo, Tokyo 113-0033, Japan\label{inst6}
            \and Department of Physics, Technion, Technion City, Haifa 3200003, Israel\label{inst7}
            \and Center for Space Sciences and Technology, University of Maryland, Baltimore County (UMBC), Baltimore, MD, 21250 USA\label{inst8}
            \and Center for Astrophysics | Harvard-Smithsonian, Cambridge, MA 02138, USA\label{inst9}
            \and Lawrence Livermore National Laboratory, Livermore, CA 94550, USA\label{inst10}
            \and Department of Astronomy, University of Michigan, Ann Arbor, MI 48109, USA\label{inst11}
            \and SRON Netherlands Institute for Space Research, Leiden, The Netherlands\label{inst12}
            \and ESO, Karl-Schwarzschild-Strasse 2, 85748, Garching bei M\"{n}chen, Germany\label{inst13}
            \and Centre for Extragalactic Astronomy, Department of Physics, University of Durham, Durham DH1 3LE, UK\label{inst14}
            \and Institute of Space and Astronautical Science (ISAS), Japan Aerospace Exploration Agency (JAXA), Kanagawa 252-5210, Japan\label{inst15}
            \and Department of Economics, Kumamoto Gakuen University, Kumamoto 862-8680 Japan\label{inst16}
            \and Department of Physics, Kyoto University, Kyoto 606-8502, Japan\label{inst17}
            \and Department of Physics, Tokyo Metropolitan University, Tokyo 192-0397, Japan\label{inst18}
            \and Department of Physics, Hiroshima University, Hiroshima 739-8526, Japan\label{inst19}
            \and Department of Physics, Fujita Health University, Aichi 470-1192, Japan\label{inst20}
            \and Department of Astronomy and Physics, Saint Mary's University, Nova Scotia B3H 3C3, Canada\label{inst21}
            \and California Institute of Technology, Pasadena, CA 91125, USA\label{inst22}
            \and European Space Agency (ESA), European Space Research and Technology Centre (ESTEC), 2200 AG Noordwijk, The Netherlands\label{inst23}
            \and Faculty of Engineering, University of Miyazaki, 1-1 Gakuen-Kibanadai-Nishi, Miyazaki, Miyazaki 889-2192, Japan\label{inst24}
            \and RIKEN Nishina Center, Saitama 351-0198, Japan\label{inst25}
            \and Leiden Observatory, University of Leiden, P.O. Box 9513, NL-2300 RA, Leiden, The Netherlands\label{inst26}
            \and Kavli Institute for Astrophysics and Space Research, Massachusetts Institute of Technology, Cambridge, MA 02139, USA\label{inst27}
            \and Department of Physics, Saitama University, Saitama 338-8570, Japan\label{inst28}
            \and Department of Physics, Rikkyo University, Tokyo 171-8501, Japan\label{inst29}
            \and Department of Physics, Tokyo University of Science, 1-3 Kagurazaka, Shinjuku-ku, Tokyo 162-8601, Japan\label{inst30}
            \and Faculty of Science and Technology, Tokyo University of Science, Chiba 278-8510, Japan\label{inst31}
            \and Department of Electronic Information Systems, Shibaura Institute of Technology, Saitama 337-8570, Japan\label{inst32}
            \and Department of Earth and Space Science, Osaka University, Osaka 560-0043, Japan\label{inst33}
            \and Department of Physics, University of Wisconsin, Madison, WI 53706, USA\label{inst34}
            \and Department of Physics \& Astronomy, Waterloo Centre for Astrophysics, University of Waterloo, Waterloo, Ontario N2L 3G1, Canada\label{inst35}
            \and Univ Toulouse, CNES, CNRS, IRAP, Toulouse, France\label{inst36}
            \and Department of Physics, Nagoya University, Aichi 464-8602, Japan\label{inst37}
            \and Science Research Education Unit, University of Teacher Education Fukuoka, Fukuoka 811-4192, Japan\label{inst38}
            \and Hiroshima Astrophysical Science Center, Hiroshima University, Hiroshima 739-8526, Japan\label{inst39}
            \and Department of Data Science, Tohoku Gakuin University, Miyagi 984-8588, Japan\label{inst40}
            \and College of Science and Engineering, Kanto Gakuin University, Kanagawa 236-8501, Japan\label{inst41}
            \and European Space Agency(ESA), European Space Astronomy Centre (ESAC), E-28692 Madrid, Spain\label{inst42}
            \and Department of Science, Faculty of Science and Engineering, KINDAI University, Osaka 577-8502, Japan\label{inst43}
            \and Department of Teacher Training and School Education, Nara University of Education, Nara 630-8528, Japan\label{inst44}
            \and Astronomical Institute, Tohoku University, Miyagi 980-8578, Japan\label{inst45}
            \and Department of Physics, Nara Women's University, Nara 630-8506, Japan\label{inst46}
            \and Department of Astrophysics and Atmospheric Sciences, Kyoto Sangyo University, Kyoto 603-8555, Japan\label{inst47}
            \and School of Science and Technology, Meiji University, Kanagawa, 214-8571, Japan\label{inst48}
            \and Yale Center for Astronomy and Astrophysics, Yale University, New Haven, CT 06520-8121, USA\label{inst49}
            \and Department of Physics, Konan University, Hyogo 658-8501, Japan\label{inst50}
            \and Graduate School of Science and Engineering, Kagoshima University, Kagoshima, 890-8580, Japan\label{inst51}
            \and Department of Physics, Chuo University, Tokyo 112-8551, Japan\label{inst52}
            \and Faculty of Education, Shizuoka University, Shizuoka 422-8529, Japan\label{inst53}
            \and Kanazawa University, Kanazawa, 920-1192 Japan\label{inst54}
            \and Department of Astronomy, Kyoto University, Kyoto 606-8502, Japan\label{inst55}
            \and Nihon Fukushi University, Shizuoka 422-8529, Japan\label{inst56}
            \and Anton Pannekoek Institute, the University of Amsterdam, Postbus 942491090 GE Amsterdam, The Netherlands\label{inst57}
            \and Department of Physics, Faculty of Science, Nara Women's University, Nara 630-8506, Japan\label{inst58}
            \and Johns Hopkins University, Baltimore, MD 21218, USA\label{inst59}
            \and Department of Astronomy and Astrophysics, University of Chicago, Chicago, IL 60637, USA\label{inst60}
            \and Department of Physics and Astronomy, The University of Georgia, Athens, GA 30602, USA\label{inst61} 
            \and D\'{e}partement de Physique, Universit\'{e} de Montréal, Succ. Centre-Ville, Montr\'{e}al, Qu\'{e}bec, H3C 3J7, Canada\label{inst62}
            \and Frontier Research Institute for Interdisciplinary Sciences, Tohoku University, 6-3 Aramakiazaaoba, Aoba-ku, Sendai, Miyagi 980-8578, Japan\label{inst63}
            \and Department of Theoretical Physics and Astrophysics, Masaryk University, Brno 61137, Czechia\label{inst64}
            \and Department of Intelligent Control and Information Engineering, National Institute of Technology (KOSEN), Kumamoto College, 2659-2 Suya, Koshi, Kumamoto 861-1102, Japan\label{inst64}
            }

   \date{Received February 26, 2026 / Accepted June 11, 2026}

  \abstract 
   {The intracluster medium (ICM) is rich in chemical elements, produced by core-collapse (SNcc) and Type Ia supernovae (SNIa) over the last $\sim$12~Gyr. While cluster outskirts are uniformly enriched with Fe at $\sim$0.3~solar, strongly suggesting that the gas had been pre-enriched during or before the assembly of galaxies into clusters, the Fe abundance is known to centrally increase in the core of relaxed clusters. The origin of these central Fe peaks however, as well as the apparent presence of mysterious drops previously reported in the very centre of a number of systems, remain to be clarified.}
   {In this paper, we address these two questions by measuring the spatial distribution of Fe and its relative Si/Fe, S/Fe, Ar/Fe, Ca/Fe, Cr/Fe, Mn/Fe, and Ni/Fe ratios in the X-ray bright, nearby Perseus cluster.} 
   {We take advantage of the unprecedented spectral resolution ($\sim$5~eV) offered by the Resolve microcalorimeter on board \XRISM, which observed four distinct pointings of Perseus out to $\sim$250~kpc ($\sim$0.2$r_{500}$) during its performance verification phase.}
   {Although the presence of an X-ray bright AGN challenges a precise quantification of absolute abundances in the very core, our baseline analysis rules out a strong drop with $>$2$\sigma$ confidence, at variance with previous charge-coupled device (CCD) measurements. In addition, we find a remarkable spatial uniformity of X/Fe ratios, supporting the idea of negligible late SNIa enrichment from the brightest cluster galaxy NGC\,1275. We also compare the overall chemical composition of the Perseus ICM with SNcc and SNIa nucleosynthesis yield models, finding that the co-existence of two separate SNIa enrichment channels is not needed to reproduce the ICM ratios satisfactorily.}
   {}


   \titlerunning{Chemical enrichment of the Perseus cluster core seen by \XRISM/Resolve}
   \authorrunning{XRISM Collaboration et al.}
   \maketitle
   \nolinenumbers

\section{Introduction}\label{sec:intro}

While the chemical composition of the early Universe primarily consisted of H and He \citep{planck2020}, the intracluster medium (ICM) is known to be polluted by Fe \citep{mitchell1976,serlemitsos1977} and many other species such as C, N, O, Ne, Mg, Si, S, Ar, Ca, Cr, Mn, and Ni \citep[e.g.][]{lea1982,mushotzky1996,deplaa2007,mernier2016}. This finding is remarkable for two reasons: first, because this hot, X-ray emitting, metal-rich plasma pervades the megaparsec-wide volume of galaxy clusters (which are known to be the largest gravitationally bound structures of our Universe); and, second, because all of these elements were once synthesised in the core of stars and/or their end-of-life explosions as supernovae \citep{burbidge1957,nomoto2013}. In other words, the presence of metals in the ICM means that clusters act as cosmic repositories of stellar end products, which have been created then dispersed out of their host galaxies over several billions of years. In addition, the simple physics governing the X-ray emission of the ICM (essentially collisional ionisation equilibrium and optically thin plasma) allows chemical abundances to be measured in clusters fairly accurately via X-ray spectroscopy, and thus to shed light on when and how this large-scale enrichment took place (for reviews, see \citealt{werner2008,biffi2018,mernier2018c}).

While direct redshift studies of cluster metallicities are feasible -- yet at limited precision with current X-ray facilities \citep[e.g.][]{mcdonald2016,mantz2017,liu2018,flores2021} -- our best clues to decipher cluster enrichment history are found in nearby systems, namely via (i) radial profiles of the (overall) metallicity and (ii) abundances of individual elements (or rather their relative X/Fe ratios), which are often estimated over the entire cluster core for better statistics. Albeit at moderate spectral resolution, the former observable revealed that, despite a radial decrease in (mostly relaxed) cluster cores, the metallicity flattens to a spatially homogeneous level of $\sim$0.3~solar out to at least $r_{200}$\footnote{This limit is defined as its delimited density corresponds to 200 times the critical density of the Universe.} \citep{fujita2008,werner2013,urban2017,ghizzardi2021}. Corroborated with cosmological hydrodynamical simulations \citep{biffi2017,biffi2018b}, this uniform metallicity floor in layers of poorly virialised gas implies that the majority of metals must have been ejected and thoroughly mixed outside their galaxy hosts more than $\sim$10~Gyr ago ($z \gtrsim 2$--3), when clusters were not (fully) assembled yet, through early feedback of active galactic nuclei (AGNs). This pre- (or early-) enrichment scenario is also in line with the second observable -- i.e. the chemical composition of the core ICM, which, besides a few possible exceptions \citep[][]{kara2024,mernier2026}, is observed to be similar from system to system \citep{deplaa2007,mernier2016,mernier2018}. In fact, this near-ubiquitous chemical composition in the central ICM may suggest that the central brightest cluster galaxy (BCG), whose properties (stellar mass, merging and star formation histories, etc.) somewhat varies from cluster to cluster, has a limited impact on post-assembly enrichment of cluster cores. High-resolution X-ray spectroscopy is particularly vital for measuring individual elemental abundances, since resolving individual lines enables a number of systematic uncertainties to be discarded in the spectral fitting \citep[e.g.][]{gastaldello2021}. In this context, measurements of the core of the Perseus cluster with the soft X-ray spectrometer (SXS) microcalorimeter instrument on board \hitomi revealed a chemical composition close to the aforementioned studies, which were all achieved at a moderate or dispersive resolution, but also surprisingly similar to the (proto-) solar composition \citep{hitomi2017,simionescu2019}.

Although they support the pre-enrichment scenario, the above two observables remain essentially indirect. While decisive light on the cosmic epoch(s) of enrichment will be revealed through accurate studies at high redshift -- and hence will have to wait for \textit{NewAthena} \citep{cucchetti2018}, a third, equally valuable angle to consider is that of the spatial variation of the ICM chemical composition. In fact, if most metals were injected and mixed thoroughly in the forming ICM early on, rather than by late outflows of cluster galaxies, one should find a fairly uniform relative contribution of Type Ia supernovae (SNIa) and core-collapse supernovae (SNcc) products in clusters, both radially and azimuthally. Using the $\sim$120~eV energy resolution of the X-ray imaging spectrometer (XIS) and the European photon imaging cameras (EPIC) on board \suzaku and \XMM, respectively, previous studies have indeed tended to suggest a radial invariance of $\alpha$/Fe ratios with radius \citep[e.g.][]{simionescu2009,simionescu2015,mernier2017,ezer2017}. This picture, however, remains under debate as hints of spatial variation in the composition of lower-mass systems have recently been reported \citep{mernier2022,sarkar2022}.

Another widely open question is that of central abundance drops. While metal peaks, ubiquitously seen in cool-core systems, extend out to at least several tens of kiloparsecs, a drop is sometimes observed within the central $\lesssim$10--20 kpc. Mostly seen in lower-mass systems \citep{panagoulia2015,mernier2017} but also in a few clusters \citep{fabian2011,sanders2016}, these metal drops are challenging to explain astrophysically. While they may be the artificial result from incorrect or over-simplistic assumptions in spectral modelling (e.g. He sedimentation, resonant scattering, temperature structure models, and atomic code uncertainties; \citealt{ettori2006,sanders2006,werner2006,fukushima2022}), another intriguing possibility is that they result from metals being depleted from the ICM phase to dust phase, and thus vanishing from the X-ray regime \citep{panagoulia2015}. Since noble gases (e.g. Ne and Ar) are much less efficiently depleted than the other elements (e.g. Fe), the latter scenario can be tested by searching for spatial variations of the Ne/Fe or Ar/Fe ratios in the central ICM. However, the issue remains unsolved as recent efforts devoted to this experiment (based again on moderate-resolution spectroscopy) have not converged yet in their conclusions \citep{lakhchaura2019,liu2019,fukushima2022}. 

Fortunately, these open questions are also accompanied by the recent launch of the \textit{X-ray Imaging and Spectroscopy Mission} (\XRISM) and its  microcalorimeter instrument Resolve. Such a transformative instrument, capable of non-dispersive high-resolution ($\sim$5~eV) spectroscopy on extended sources such as clusters, allows to resolve emission lines from Si to Ni (in its closed gate valve configuration) and to map spatially their abundances with higher confidence than previous X-ray spectrometers. This is particularly relevant for bright, nearby systems on which this mapping can be achieved reliably, even despite the limited point spread function (PSF) of the telescopes ($\sim$1.2 arcmin at half-power diameter). Among them, Perseus -- the brightest X-ray cluster in the sky, is a target of prime choice. This extremely nearby system is also cool-core, with an ideally moderate ICM temperature ($\sim$ 3--7~keV) to isolate emission lines and continuum accurately. Last but certainly not least, Perseus is known to host a central abundance peak with an apparent drop in its central $\sim$50 kpc \citep[e.g.][]{sanders2004,sanders2007,fabian2011,mernier2017}. Although the Perseus core had already been observed by \hitomi \citep[e.g.][]{hitomi2016,hitomi2017,hitomi2018temp,hitomi2018vel,simionescu2019}, the SXS pointings were directed almost exclusively towards the very core, with very little information on the spatial variation of its ICM abundances.

In this paper, we take advantage of the first \XRISM/Resolve observations of the core of the Perseus cluster to measure chemical abundances -- and their spatial distributions -- with unprecedented accuracy. The four spatially separated pointings considered here, taken during the performance verification (PV) phase of the mission, allow us to address the two key questions mentioned above: (i) the spatial variation (or uniformity) of the chemical composition of the Perseus central ICM and (ii) the existence (and characterisation) of its central metal drop. This work follows a first paper, based on the same observations but dedicated to its gas dynamics, which is found to be driven by both AGN feedback (small-scale motions) and merger-related processes (large-scale motions) in the cool core \citep[][hereafter PaperVel]{xrism2025_Perseus}. Sections~\ref{sec:reduction} and ~\ref{sec:spectra} describe respectively the data reduction (including our mapping strategy) and our spectral fitting approach (including the treatment of PSF effects). Our measurements are presented in Sect.~\ref{sec:results} and are then interpreted in Sect.~\ref{sec:discussion}. Finally, Sect.~\ref{sec:conclusion} concludes our study.
We assume a standard $\Lambda$CDM cosmology with $H_0 = 70$~km\,s$^{-1}$\,Mpc$^{-1}$, $\Omega_\Lambda = 0.73$ and $\Omega_m = 0.27$. At a redshift of $z = 0.017284$ for Perseus' BCG, NGC\,1275 \citep{hitomi2018vel}, 1~arcmin corresponds approximately to 22~kpc. Abundances derived in this paper are expressed in proto-solar units of \citet{lodders2009}. Error bars are given at a 68\% confidence level.

\section{Data reduction}\label{sec:reduction}

\begin{figure}[t!]
\centering
\includegraphics[width=0.45\textwidth, trim={1.5cm 0cm 1.5cm 7cm},clip]{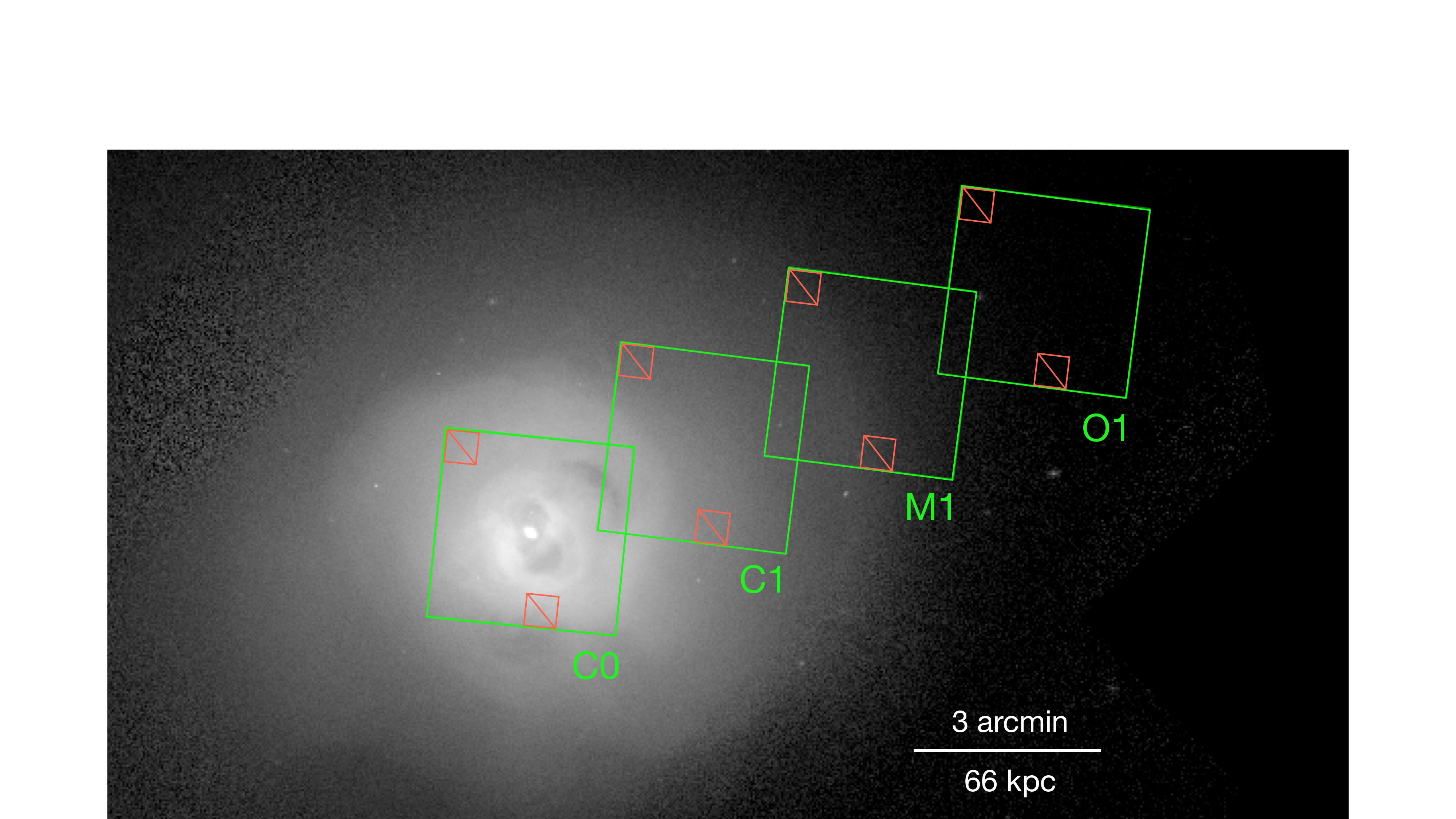} 
\includegraphics[width=0.45\textwidth, trim={1.5cm 0cm 1.5cm 7cm},clip]{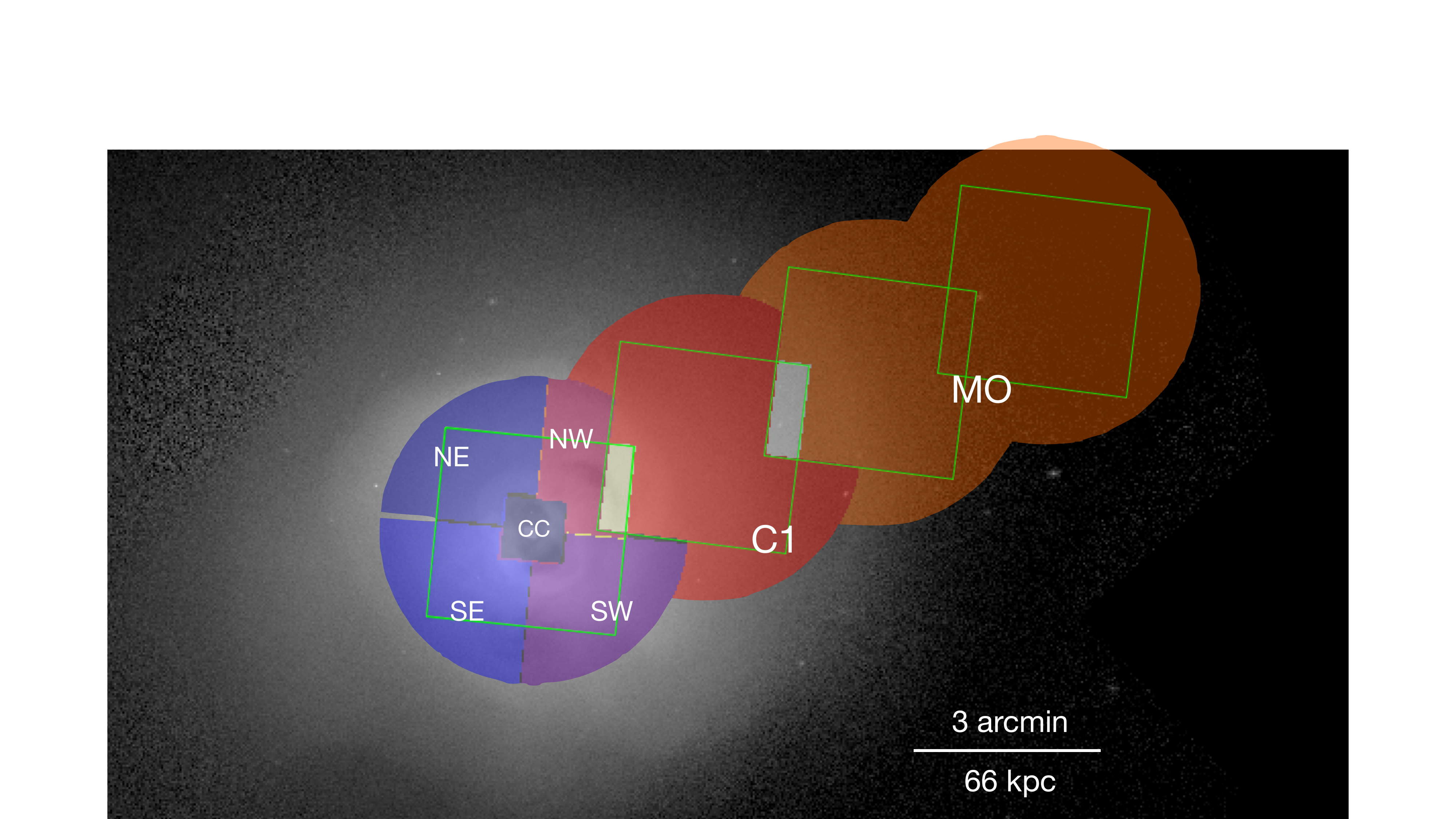} 
      \caption{Mosaiced \chandra/ACIS image (1.8--9~keV) of the core of the Perseus cluster superimposed with our Resolve regions of interest. \textit{Top:} The four full-array Resolve pointings (green) excluding pixels 12 and 27 (red). \textit{Bottom:} Our seven sky mapping regions including Resolve pixels, selected to extend beyond the Resolve array boundaries when appropriate (see text). The grey areas mark an overlap between two regions (NW-C1 and C1-MO; see text). }
         \label{fig:map}
\end{figure}

During its PV phase, \textit{XRISM} observed the core of Perseus across four pointings: C0, C1, M1, and O1 (Table~\ref{table:obslog}, Fig.~\ref{fig:map}). The C0 pointing, which covers the X-ray bright AGN of NGC\,1275 as well as the innermost ICM cavities, consists of two separate observations which are treated independently throughout this work.

\begin{table}[h!]
\caption{Summary of the \XRISM observations of Perseus analysed in this paper.}                 
\label{table:obslog}    
\centering 
\begin{tabular}{l | c c c} 
\hline\hline 
Pointing & ObsID & Observation date & Net exposure \\         
 &  &  & (ks) \\         
\hline          

C0 (\#1) & 000154000 & 2024-01-21 & 48.3 \\
C0 (\#2) & 000155000 & 2024-01-22 & 50.0 \\
C1 & 000156000 & 2024-01-23 & 57.1 \\
M1 & 000157000 & 2024-01-24 & 92.6 \\
O1 & 000158000 & 2024-01-26 & 131.2 \\

\hline                                  
\end{tabular}
\end{table}

\subsection{Event lists, spectra, and responses}\label{sec:reduction:basic}

The first-step reduction of the Resolve data followed essentially the standard recommendations from the public Data Reduction Guide\footnote{\href{https://heasarc.gsfc.nasa.gov/docs/xrism/analysis/}{https://heasarc.gsfc.nasa.gov/docs/xrism/analysis/}} and the previous Resolve study of Perseus gas dynamics \textcolor{blue}{(PaperVel)}. We used the Build 8 version of the \textit{XRISM} data reduction software with the calibration database CalDB v8, both of which give results that are virtually identical\footnote{We have specifically verified over the C1 pointing that the results discussed here are virtually identical to those obtained with HEASoft v34, CalDB v10.} to the results of the first public version available in HEASoft\footnote{\href{https://heasarc.gsfc.nasa.gov/docs/software/lheasoft/}{https://heasarc.gsfc.nasa.gov/docs/software/lheasoft/}} and CalDB v10. After preprocessing the data using the command \texttt{xapipeline}, we applied an extra rise-time screening to minimise the background. We then extracted Resolve spectra for each region of interest (see Sect.~\ref{sec:reduction:region} for a complete description of these) after having ignored pixel 12 -- used for calibration outside the field of view (FoV) -- and pixel 27 -- showing hardly predictable gain excursions at the time of the observations. For each spectrum, we used the command \texttt{rslmkrmf} to create an `extra-large' redistribution matrix file (RMF), which accounts for the entire shape of the line spread function as well as the electron loss continuum below $\lesssim3$~keV. We made sure to use the up-to-date RMF parameter file (\texttt{xa\_rsl\_rmfparam\_20190101v006.fits}; see also \citealt{leutenegger2025}). The ancillary response files (ARF) were generated via the command \texttt{xaarfgen}, which includes a ray-tracing tool that calculates the fraction of photons coming from a certain sky region (e.g. an input high-resolution image) into the `detector' region over which a given spectrum is extracted \citep[\texttt{xrtraytrace};][]{yaqoob2018}. These ARFs were systematically generated in `image mode', based on a \textit{Chandra}/ACIS image of Perseus in the 1.8--9~keV band. In addition, we generated an extra set of ARFs in `point-source' mode to account for the  point-like nature of central AGN component (Sect.~\ref{sec:spectra:components}). A detailed use of these ARFs, adapted to our science case, is developed in Sect.~\ref{sec:spectra:SSM}.

 \subsection{Regions of interest}\label{sec:reduction:region}

Our efficiency to derive Resolve spatial abundance maps of the Perseus core is dictated by two constraints: (i) the count statistics of each extracted region and (ii) the $\sim$1.2 arcmin extent (half-energy width) of the PSF. This implies a typical size for our spatial regions, which ought to be neither too small nor too large. Here, we chose to define regions such that they include at least 40,000 Resolve counts, which is necessary to achieve $\sim$10--20\% precision on the abundance of most elements (see Sect.~\ref{sec:results}). While the C1 pointing meets this criterion by itself, the M1 and O1 pointings need to be combined (`MO' region; via a coupled modelling of the M1 and O1 spectral components) to reach sufficient statistics. On the other hand, the C0 pointing can be split in four azimuthal regions (`SE', `NE', `SW', and `NW') as well as one region `CC', consisting of the central four pixels, which covers a mixed contribution of the emission of the central AGN and the surrounding ICM\footnote{The CC region contains approximately 57,000 Resolve counts, which exceed our 40,000 criterion. However, calibration uncertainties and the size of the PSF do not allow to subdivide this region any further.}. 

These seven regions are showcased in Fig.~\ref{fig:map} (bottom panel). We defined them to extend out to a maximum distance $r_\mathrm{max} \le 2.5$~arcmin from the centre of their parent Resolve pointing to account for a (small but non-negligible) contribution from photons coming from outside the detector, yet falling into the considered Resolve pixels (see also \citealt{hitomi2018vel} and Sect.~\ref{sec:spectra:SSM}). In addition, we note an overlap of pixels between NW and C1 on one hand, and between C1 and MO on the other hand. Instead of stacking these overlapping corners into a dedicated region, we chose to treat them separately in their individual regions -- thus to keep the overlap as it is. Doing so ensures that we avoid systematic complications in the spectral fits, for instance due to residual cross-calibration issues between pixels from different locations on the detector.

\section{Spectral fitting}\label{sec:spectra}

\subsection{Spectral components}\label{sec:spectra:components}

Throughout this paper, the spectral fitting was done using the \texttt{Xspec} fitting package embedded with AtomDB v3.0.9\footnote{We have verified that our results are not affected by the use of the recently updated AtomDB v3.1.3 version.}. Unless otherwise stated, our Resolve spectra were fitted within $E_\mathrm{fit} \in [1.9, 12]$~keV. Seen at high spectral resolution, resonant scattering is known to affect the Fe\,{\sc xxv} resonant (w) line of cluster cores, specifically in Perseus were this line is seen to be suppressed by a factor $\sim$1.3 in the inner 30~kpc \citep{hitomi2018RS}. To ensure this suppression does not impact our Fe abundance measurement, we excluded this line from all our fits (i.e. $E_\mathrm{fit} \notin [6.567, 6.620]$~keV)\footnote{An alternative method, fitting the Fe\,{\sc xxv} w line separately with a Gaussian function and giving results that are consistent with ours, is presented in \textcolor{blue}{PaperVel}. Moreover, we note that treating this line as optically thin in our fits does not significantly alter our absolute Fe abundance measurements and discussions addressed in Sects.~\ref{sec:results:Fe} and \ref{sec:discussion:drop}.}. A complete investigation of resonant scattering in our dataset will be published separately (\textcolor{blue}{Heinrich et al., in prep.; hereafter PaperRS}). Our fitting method is based on the C-statistics \citep{kaastra2017} and we ensured that each spectral bin contains at least one count. The three spectral components to model are (i) the ICM emission, (ii) the central AGN emission (where relevant) and (iii) the non-X-ray background (NXB). We note that the astrophysical foreground, originating from the Local Hot Bubble and the Milky Way hot halo, emits mostly below $E \lesssim 1$~keV and can thus be largely ignored in the closed gate valve configuration.

The ICM component was modelled by a thermal plasma of the \texttt{Xspec} `\texttt{bvv}'-family (i.e. with variable velocity broadening and abundance of all $6 \le Z \le 30$ elements) absorbed by a neutral hydrogen column density model (\texttt{TBabs}) with $n_\mathrm{H} = 13.6 \times 10^{20}$~cm$^{-2}$ \citep{willingale2013}. Although the literature contains a number of examples with only one thermal component, cool-core clusters -- and in particular Perseus \citep{hitomi2018temp} -- are known for being multi-temperature at least in projection. For this reason, which will be further elaborated in Sect.~\ref{sec:results:systematics}, this work is based on several (single- or multi-temperature) modelling strategies as listed below.

\begin{itemize}
	\item 1T, a single-temperature model (\texttt{TBabs*bvvapec}) with the normalisation, temperature, redshift, velocity dispersion and Si, S, Ar, Ca, Cr, Mn, Fe, and Ni abundances left free. The other abundance parameters were tied to that of Fe.
	\item 2T, a two-temperature model (\texttt{TBabs*(bvvapec} \texttt{+bvvapec)}) with the two normalisations, temperatures, redshifts, and velocity dispersions left free and independent from each other. To avoid overfitting, the abundances of the two components were tied with each other.
	\item gT, a multi-temperature model with differential emission measure shaped as a Gaussian distribution (\texttt{TBabs*bvvgadem}) with its temperature width $\sigma_T$ left free. The other parameters (normalisation, redshift, velocity dispersion, abundances) follow the 1T prescription. For reasons developed in Sect.~\ref{sec:results:systematics}, this choice constitutes our baseline model.
	\item wT, a multi-temperature model with differential emission measure shaped as a power law (\texttt{TBabs*bvvwdem}). Namely, its distribution follows $\propto kT^{1/\alpha_T}$ with the (inverse) slope parameter $\alpha_T$ treated as a free parameter. The above distribution holds for temperatures situated between $\beta_T kT_\mathrm{W} < kT < kT_\mathrm{W}$, where the parameter $kT_\mathrm{W}$ (representing the maximum temperature of the distribution) was left free and the temperature ratio $\beta_T$ (used to construct the lower limit of the distribution) was fixed to its default value of 0.1. Outside this interval, the emission measure is zero. The other parameters (normalisation, redshift, velocity dispersion, abundances) follow the 1T prescription.
\end{itemize}

Although the central AGN of NGC\,1275 was extensively studied in the past \citep[e.g.][]{hitomi2018AGN,rani2018,reynolds2021}, the source is known to be variable and related priors should be established with caution. Following \textcolor{blue}{PaperVel}, we assumed in the CC region an absorbed power-law continuum and a Gaussian (fluorescent) line at 6.4~keV with free normalisations. Our baseline approach allows the photon index to vary within $\Gamma_\mathrm{AGN} \in [1.6, 1.9]$ -- consistently with the range constrained in that earlier work -- and the normalisation to vary freely. This approach is further revisited in Sect.~\ref{sec:results:systematics} and we note that a specific Resolve study to characterise this AGN is ongoing \textcolor{blue}{(Ogorzalek et al. in prep.; hereafter PaperAGN)}.

Last but not least, we included an NXB component in all our spectra by using a model pre-fitted on a collection of actual NXB observations. Developed by the \textit{XRISM} team\footnote{\href{https://heasarc.gsfc.nasa.gov/docs/xrism/analysis/nxb/nxb_spectral_models.html}{https://heasarc.gsfc.nasa.gov/docs/xrism/analysis/nxb /nxb\_spectral\_models.html}}, this model consists of a continuum plus a series of fluorescent lines and is controlled by three scale factors set as free parameters in each individual fit. Since these events are particles instead of photons, the NXB component was not folded by any ARF and was used with a diagonal RMF (provided separately). 

\begin{figure*}[h!]
\centering
\includegraphics[width=0.89\textwidth, trim={0.0cm 0cm 0.0cm 0cm},clip]{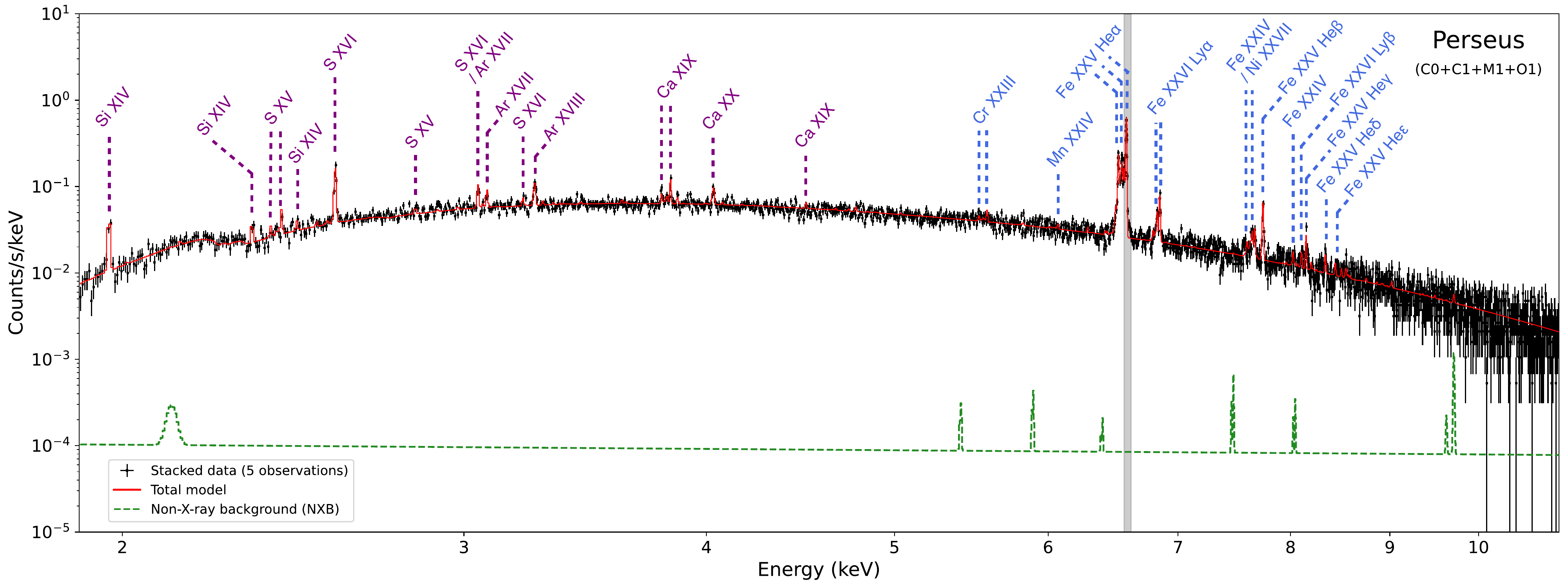} 
      \caption{Resolve spectrum stacked over the five observations (four pointings) considered in this work, for illustration purposes only. The main emission lines are shown in purple and blue, respectively for elements originating either (partly) from SNcc or (almost entirely) from SNIa. The grey band around the Fe\,{\sc xxv}~w line is not considered in our fits (see text). On average, the NXB model (dashed green) is largely sub-dominant.}
         \label{fig:spectrum}
\end{figure*}

To highlight the total statistics in hand (also relevant for Sect.~\ref{sec:discussion:composition}) and the accessible emission lines, we show in Fig.~\ref{fig:spectrum} a Resolve spectrum stacked over the four pointings (five observations including the duplicate C0; see Table~\ref{table:obslog}) along with its combined total best-fit and NXB models. We stress that this figure is for illustration purpose only and does not represent a direct fitting: it is instead obtained by combining separately the observed spectra and their corresponding best-fit models from each pointing before superimposing them on the figure. Relevant best-fit abundance ratio parameters obtained over all regions are presented further in Sect.~\ref{sec:discussion:composition}. Although the NXB component is seen to be marginal overall, it can reach approximately 10\% (or more near the fitted energy limits) in the O1 pointing. It is therefore important to keep a careful NXB modelling into consideration in our analysis.

\subsection{Treatment of the spatial-spectral mixing}\label{sec:spectra:SSM}

As stated earlier, the extent of the \XRISM PSF (i.e. about 40\% of the Resolve FoV) means that a non-negligible fraction of photons observed in a given region originates from elsewhere. This spatial-spectral mixing (SSM) effect needs to be accounted for properly in the spectral modelling. Concretely, doing spectral analysis of neighbouring regions (with the aim of obtaining a spatial map of a given observable) requires the spectrum of each region to be cross-modelled by components describing the other regions, and vice-versa. 

A comprehensive description of this method is provided in \citet{hitomi2018vel} and can be summarised by separating a given `sky' region $K$ -- corresponding to the true region one wants to investigate -- from its `detector' region counterpart $k$ -- corresponding to the region actually observed (and contaminated by SSM). Specifically, a spectrum $S_i$ observed in detector region $i$ can be modelled by a linear combination
\begin{equation}
    S_i = R_i \ast \sum_J A_{J\rightarrow i} \ast M_J ,
\end{equation}
where $R_i$ is the RMF extracted from the same detector region $i$, $M_J$ is the spectral model describing sky region $J$ (which may or may not overlap with detector region $i$) and, last but not least, $A_{J\rightarrow i}$ is the ARF computed exclusively for photons from (sky) region $J$ leaking into (detector) region $i$. During its generation, the ARF is thus normalised to the right fraction of leaking photons thanks to \texttt{xrtraytrace} within the \texttt{xaarfgen} command. These fractions $f_{J\rightarrow i}$ are listed in Table~\ref{table:SSMcoef} (Appendix~\ref{sec:app:SSMcoef}). Since our mapping scheme has seven regions, this method implies in principle to gather our seven spectra in one joint fit containing $7 \times 7 = 49$ \texttt{TBabs*bvgadem+NXB} components, with all parameters coupled appropriately. This setup becomes even heavier considering that (i) the SE, NE, CC, SW, and NW spectra are in fact duplicated as they belong to two separate ObsIDs (000154000 and 000155000); (ii) the CC region (and its spectral component to be modelled in the neighbouring regions) has the AGN component described above, and (iii) the MO region is actually made of two pointings, each having their own thermal and NXB components. To ease the analysis, we consider that the fraction $f_{J\rightarrow i}$ of photons from region $J$ leaking into region $i$ is negligible when $i$ and $J$ are not adjacent. For instance, we find that $f_{CC\rightarrow C1}$ and $f_{NE\rightarrow SW}$ amount to 1.5\% and  4.3\%, respectively. In total, we were left with a master fit of 12 spectra, modelled altogether with a total of 76 components (i.e. 53 ICM components, 10 AGN components, and 13 NXB components).

\section{Results}\label{sec:results}

\subsection{Uncertainties from spectral modelling}\label{sec:results:systematics}

In principle, abundance measurements can be biased by a number of effects. The three most important ones, following lessons learned from CCD-resolution studies over the last decades, are: 
\begin{enumerate}
    \item An over-simplified treatment of the ICM (multi-) temperature structure, which may lead to a mismodelling of the unresolved Fe-L complex (relevant below 2 keV and at moderate resolution) but also to an incorrect conversion of the line equivalent width into abundance (relevant for this work);
    \item The presence of the X-ray bright central AGN in Perseus, whose power-law continuum mixes with the continuum of the ICM emission;
    \item Background emission, also impacting the measured ICM continuum when its components are non-negligible.
\end{enumerate}

The stable and low NXB of Resolve (due to the low-Earth orbit of the mission), considerably mitigates background as a source of uncertainty. Although it is never negligible by itself (thus needs to be accounted for in all our spectra), the NXB is modelled in a flexible way (Sect.~\ref{sec:spectra:components}) which accounts for residual deviations from the initial expectations. Moreover, the Perseus ICM remains dominant by at least one order of magnitude in all the investigated regions. While the main sources of foreground (i.e. the Local Hot Bubble and Milky Way diffuse emission) are invisible beyond 2~keV, we also verified that neither the unresolved cosmic X-ray background (CXB) nor resolved point sources pollute our pointings significantly. Specifically, in the 2-10 keV band the CXB and resolved point sources account for less than, respectively, $\sim$1.5\% and $\sim$0.5\% in all considered regions. We are thus left with the first two effects which should be treated with caution in our case.

\begin{figure}[h!]
\centering
\includegraphics[width=0.49\textwidth, trim={1.0cm 0cm 1.5cm 0cm},clip]{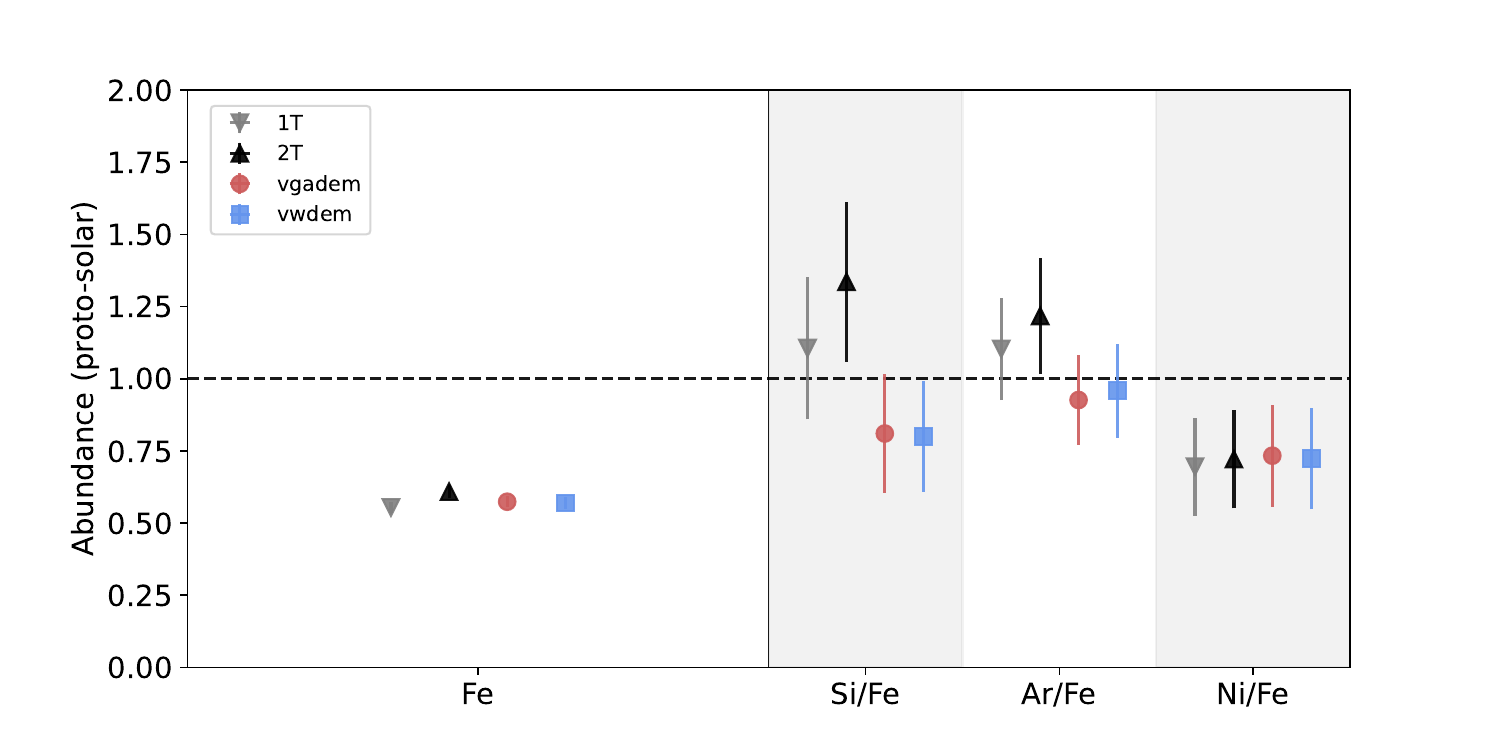} 
      \caption{Fe dependency on multi-temperature models in the C1 pointing (full array). For simplicity, no SSM is considered.}
         \label{fig:multiT}
\end{figure}

To verify the impact of the temperature structure modelling on the abundances, we perform a series of fits assuming successively the single- and multi-temperature ICM models as described in Sect.~\ref{sec:spectra:components}. We choose to compare these fits in the (full-array) C1 pointing, selected for its high count statistics, its limited SSM from the very core, and its absence of AGN contamination. The Fe abundances measured in this experiment are shown in the left panel of Fig.~\ref{fig:multiT}. Encouragingly, they all show a remarkable degree of consistency. Similarly, the right panel of Fig.~\ref{fig:multiT}, shows that the Si/Fe, Ar/Fe and Ni/Fe ratios (selected for lines of their elements to span over the entire Resolve energy band) are 1$\sigma$ consistent between virtually all models. We note a trend of somewhat higher Si/Fe and Ar/Fe best-fit values when using a 1T or a 2T model, which we attribute to a slight underestimate of the softer part of the continuum (where Si and Ar lines are found), likely due to the limited number of components in these two models. We also verify that the other X/Fe ratios (not plotted here for clarity) are equally unaffected by the choice of temperature modelling. Interestingly, similar abundance consistencies across temperature modelling were found in the core of the Centaurus cluster \citep{mernier2026}. Given these results, and since Perseus is known to host a multi-temperature structure \citep[][]{hitomi2018temp}, we select gT as our baseline model to be used throughout this paper. Its advantage over the 2T and wT models resides in a lower number of modelling parameters ($kT_\mathrm{mean}$ and $\sigma_T$), which simplifies our analysis further. We highlight that a complete multi-temperature study of the Perseus core as observed with Resolve is presented in \textcolor{blue}{Meunier et al. (in prep.; hereafter PaperT)}.

As a second step, we investigate the effect of the central AGN contribution on our abundance measurements. Admittedly, a realistic modelling of the AGN component is rendered challenging given the difficulty of disentangling its continuum from that of the ICM. While this issue (and a characterisation of the central AGN emission) is addressed in detail in \textcolor{blue}{PaperAGN}, in this work we use a conservative approach based on a series of fitting experiments in the CC region, as described in Table~\ref{table:experiments}. Experiments A and B constitute our fitting baselines, respectively with and without SSM (Sect.~\ref{sec:spectra:SSM}), and assume the slope of the AGN power-law to be $1.6 \le \Gamma_\mathrm{AGN} \le 1.9$ as previously estimated in \textcolor{blue}{PaperVel}. However, a deeper investigation from \textcolor{blue}{PaperAGN} suggests somewhat steeper slopes, typically with 3$\sigma$ limits of [2.143, 2.309]. While uncertainties on the slope is known to have very limited effects on velocity estimates, they propagate on the thermal (ICM) continuum. Since line equivalent widths directly depend on the latter, AGN slope uncertainties propagate on absolute abundance estimates too. To quantify this effect, Experiments C--G explore various assumptions on $\Gamma_\mathrm{AGN}$, which include fixed values (1.7, 1.9, 2.1, 2.3) as well as thawing the parameter with no prior limits. Experiment H repeats the latter setup, but with a fixed AGN flux $f_\mathrm{AGN}$ in the 2--10~keV band to the value of $31 \times 10^{12}$~erg\,s$^{-1}$\,cm$^{-2}$ as obtained in \textcolor{blue}{PaperVel}. Finally, Experiments I--K push the lower energy limit in the fit down to 1.7~keV, assuming the two $\Gamma_\mathrm{AGN}$ ranges reported above, as well as a free $n_\mathrm{H}$ within $[10, 20] \times 10^{20}$~cm$^{-2}$ to explore uncertainties related to the hydrogen absorption along the line of sight.

\begin{table*}[h!]
\caption{Detailed description of our fitting experiments in the CC region (see text).}                 
\label{table:experiments}    
\centering                        
\begin{tabular}{l c c c c c}      
\hline\hline               
Exp.	&	SSM?	&	$E_\mathrm{fit}$	&  $f_\mathrm{AGN}$  & $\Gamma_\mathrm{AGN}$   &  $n_\mathrm{H}$ \\         
	&	 &	(keV)	&    ($10^{12}$~erg\,s$^{-1}$\,cm$^{-2}$)   &     &  ($10^{20}$~cm$^{-2}$) \\         
\hline                      
A	&	No	&	[1.9, 12]   &	free ($40.6 \pm 1.4$) &   free within [1.6, 1.9] & 13.6\\
B	&	Yes	&	[1.9, 12]   &	free ($41.1 \pm 1.3$) &   free within [1.6, 1.9] & 13.6\\
C	&	No	&	[1.9, 12]   &	free ($14.5 \pm 0.6$) &   1.7 & 13.6\\
D	&	No	&	[1.9, 12]   &	free ($24.8 \pm 0.9$) &   1.9 & 13.6\\
E	&	No	&	[1.9, 12]   &	free ($44.6 \pm 1.3$) &   2.1 & 13.6\\
F	&	No	&	[1.9, 12]   &	free ($67.4 \pm 0.4$) &   2.3 & 13.6\\
G	&	No	&	[1.9, 12]   &	free ($46 \pm 5$) &   free ($2.110 \pm 0.016$) & 13.6\\
H	&	No	&	[1.9, 12]   &	31 &   free ($1.77 \pm 0.04$)   & 13.6\\
I	&	No	&	[1.7, 12]   &	free ($24.9 \pm 0.9$) &   free within [1.6, 1.9] & 13.6\\
J	&	No	&	[1.7, 12]   &	free ($51.18 \pm 0.24$) &   free within [2.143, 2.309] & 13.6\\
K	&	No	&	[1.7, 12]   &	free ($24.6 \pm 0.9$) &   free within [1.6, 1.9] & free within [10, 20]\\

\hline                                  
\end{tabular}
\tablefoot{$f_\mathrm{AGN}$ is estimated between 2--10~keV. Best-fit $\Gamma_\mathrm{AGN}$ values that are left free within [1.6, 1.9] and [2.143, 2.309] systematically reach their upper and lower allowed limit, respectively.}
\end{table*}

\begin{figure}[h!]
\centering
\includegraphics[width=0.49\textwidth, trim={1.5cm 0cm 1.5cm 0cm},clip]{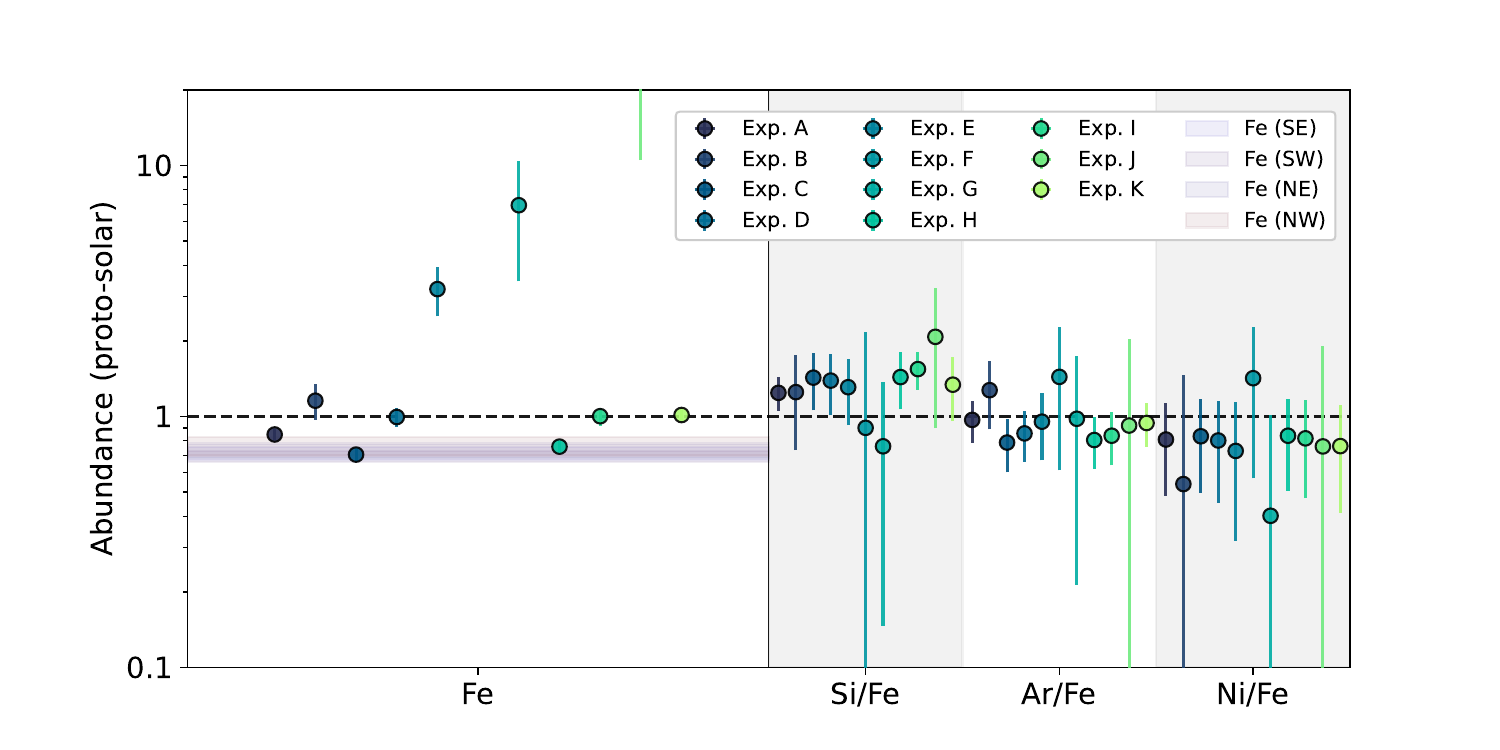} 
      \caption{\textit{Left:} Absolute Fe abundance dependencies on eleven fitting experiments in the CC region (Table~\ref{table:experiments}). The Fe values of Experiments F and J exceed the upper end of the figure and are probably not physical (see text). Fe abundance ranges measured in the surrounding SE, NE, SW, and NW regions (with SSM, i.e. following Experiment B) are indicated by light purple areas (see also Fig.~\ref{fig:Fe}). \textit{Right:} Si/Fe, Ar/Fe, and Ni/Fe dependencies on the same experiments in the CC region.}
         \label{fig:experiments}
\end{figure}

The estimates on the absolute Fe abundance and on the X/Fe ratios resulting from each experiment in the CC region are plotted in, respectively, the left and right panels of Fig.~\ref{fig:experiments}. Quite expectedly, the choice of the experiment has a strong impact on the measured absolute Fe abundance. Experiments (E,G) and (F,J) noticeably favour a best-fit Fe value larger than 3 and 10 solar, respectively. Such measurements were never reported in the core of other systems and would be hardly explainable with standard cluster enrichment scenarios. The seven other experiments report more realistic values, falling typically within 0.66--1.35~solar (including their 1$\sigma$ errors). All in all, our analysis indicates that AGN modelling uncertainties make the Fe abundance in the CC region highly uncertain, although none of our measurements are reported to be lower than $\sim$0.66~solar. This value can thus safely be considered as a lower limit in this centremost region. Much more encouragingly, AGN modelling uncertainties weakly affect our measured X/Fe ratios, as the latter remain essentially unchanged across our eleven modelling experiments (Fig.~\ref{fig:experiments}, right panel). We thus adopt Experiment B as our baseline analysis.

\subsection{Maps of absolute Fe abundance and X/Fe ratios}\label{sec:results:Fe}

\begin{figure}[h!]
\centering
\includegraphics[width=0.49\textwidth, trim={1.5cm 0cm 1.5cm 0cm},clip]{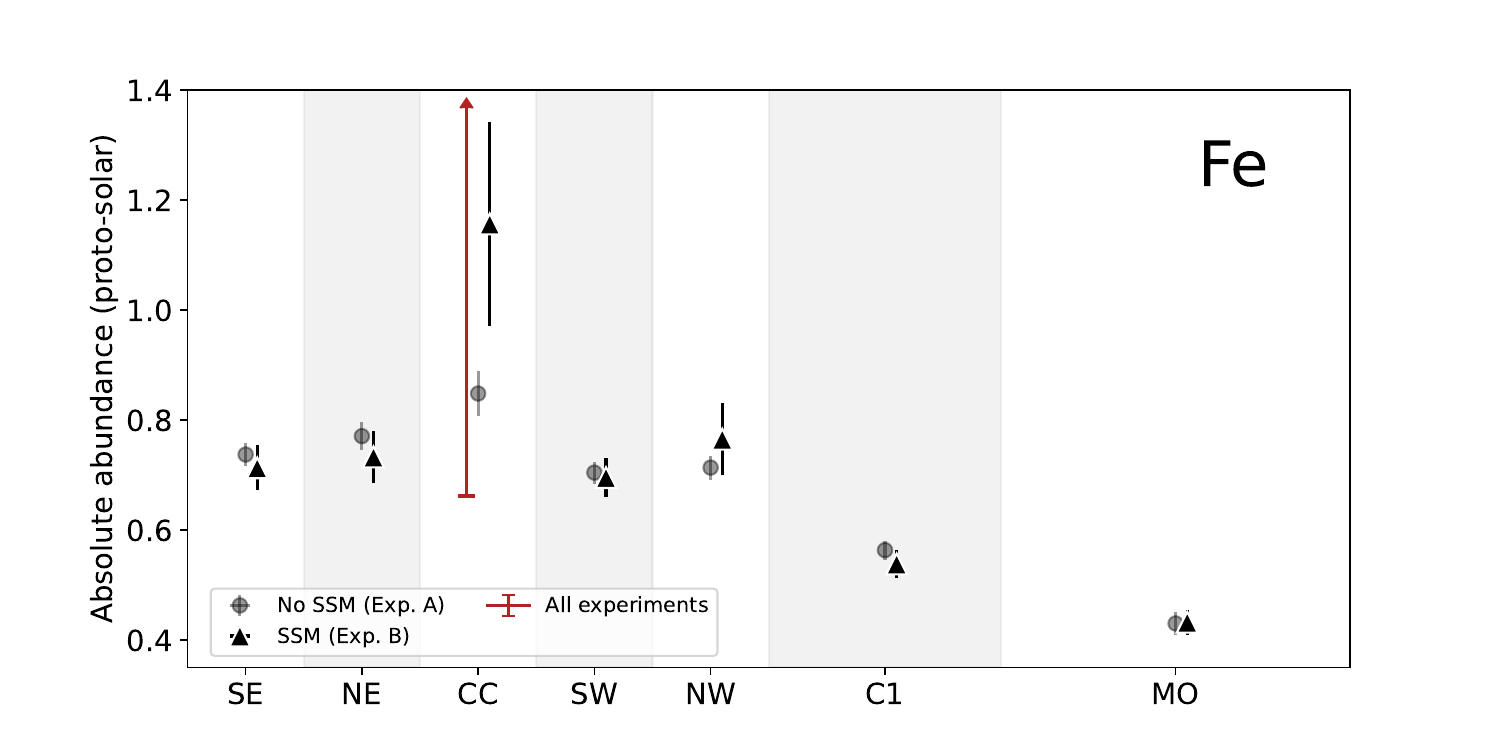} 
      \caption{Absolute Fe abundance mapped in all considered regions without and with modelling of the SSM (Experiments A and B, respectively, of our baseline analysis). The upper limit of the CC region (dark red) reflects the large dependency of Fe with our fitting experiments as illustrated in Fig.~\ref{fig:experiments}.}
         \label{fig:Fe}
\end{figure}

Considering now all regions, we show our Fe mapping results in Fig.~\ref{fig:Fe}, successively ignoring then including SSM in our analysis. This corresponds to, respectively, Experiments A and B extended to the entire map. Values of the SSM analysis are also provided in Table~\ref{table:abundances} as our final results. We note that the `SSM' Fe value of the CC region is in fact $>$1~solar, thus considerably higher than those measured in the SE, NE, SW, and NW regions. Motivated by the results from the previous section, we set an extra lower limit of $\mathrm{Fe} \ge 0.66$~solar in the CC region.

Outside the CC region, we find an excellent consistency between our measurements with and without SSM. This is encouraging, as this indicates that SSM effects are quite limited in our case, both for absolute abundances and for their X/Fe ratios (see below). Other best-fit parameters, detailed in dedicated papers (e.g. \textcolor{blue}{PaperVel}, \textcolor{blue}{PaperT}), are also minimally affected by SSM. More importantly, it clearly appears that Resolve measurements favour either a central increase or flattening of Fe. A central Fe drop, on the other hand, appears much less likely. This apparent absence of central Fe drop is discussed in Sect.~\ref{sec:discussion:drop}.

\begin{figure*}[h!]
\centering
\includegraphics[width=0.32\textwidth, trim={1.5cm 0cm 1.5cm 0cm},clip]{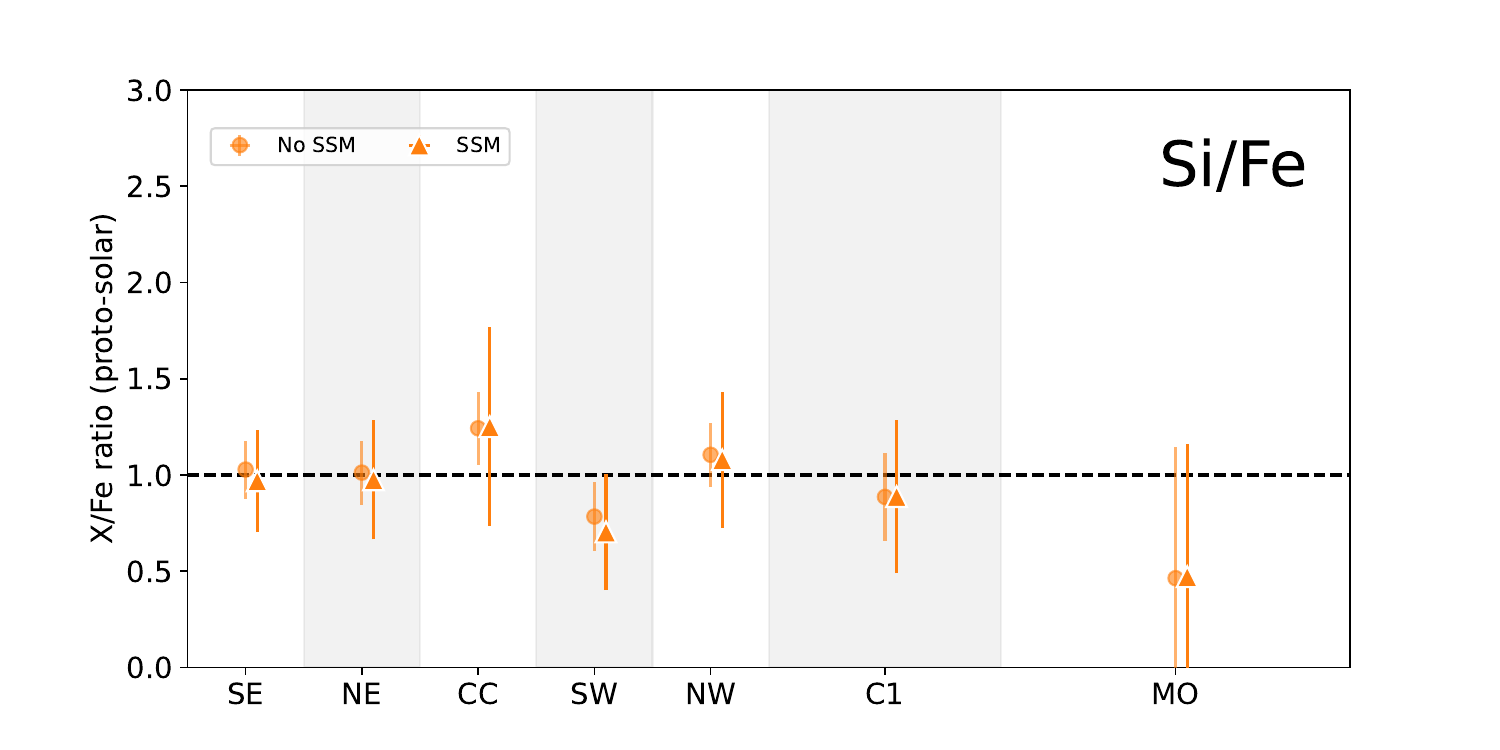} 
\includegraphics[width=0.32\textwidth, trim={1.5cm 0cm 1.5cm 0cm},clip]{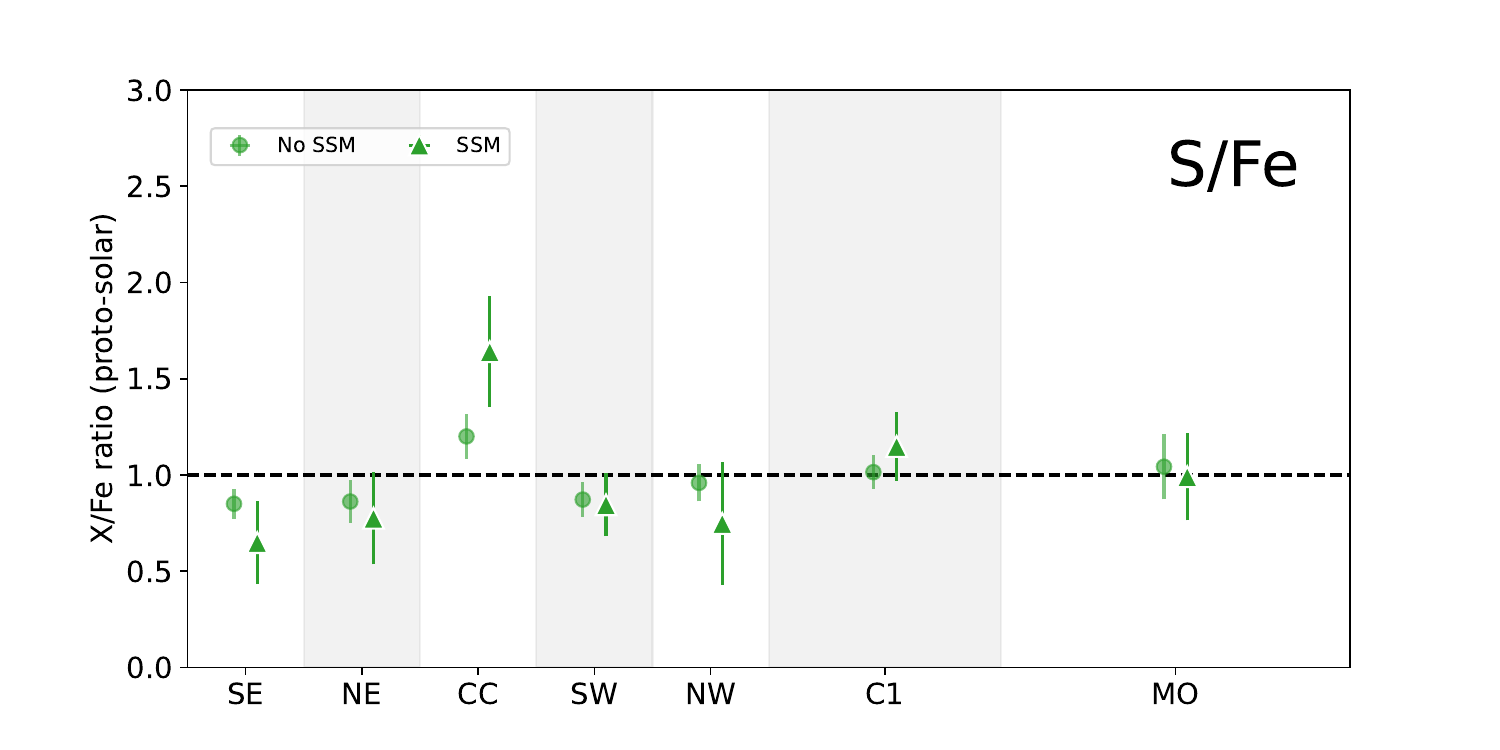} 
\includegraphics[width=0.32\textwidth, trim={1.5cm 0cm 1.5cm 0cm},clip]{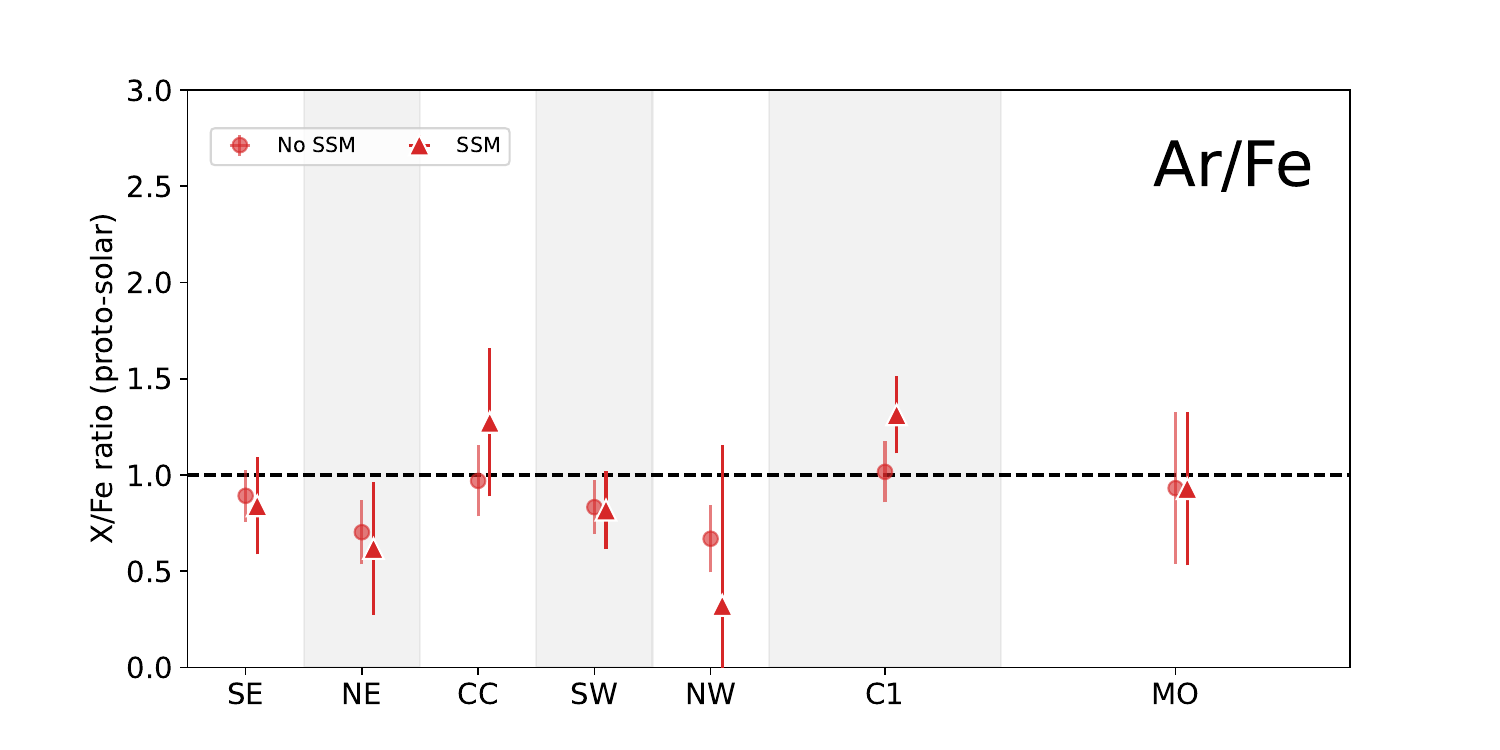} \\
\includegraphics[width=0.32\textwidth, trim={1.5cm 0cm 1.5cm 0cm},clip]{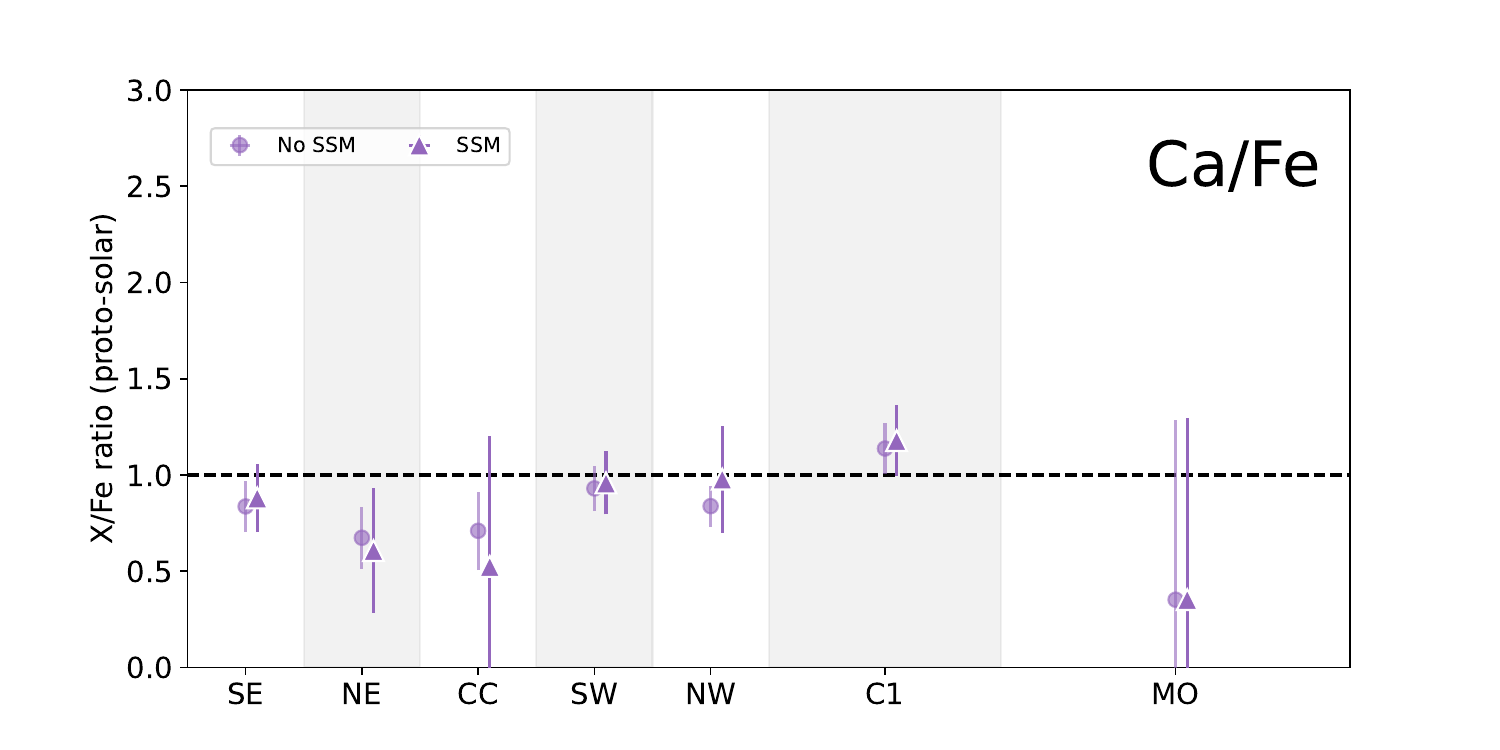} 
\includegraphics[width=0.32\textwidth, trim={1.5cm 0cm 1.5cm 0cm},clip]{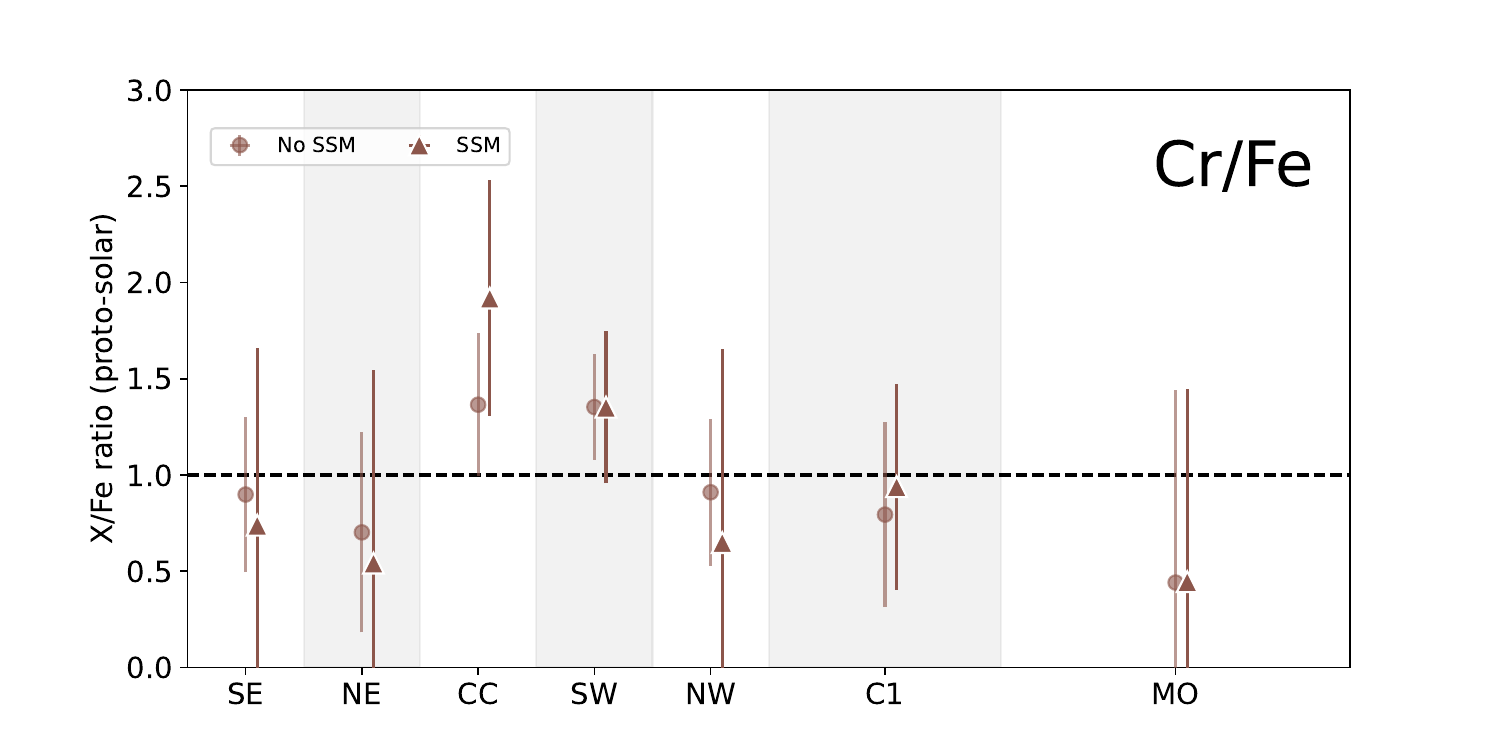} 
\includegraphics[width=0.32\textwidth, trim={1.5cm 0cm 1.5cm 0cm},clip]{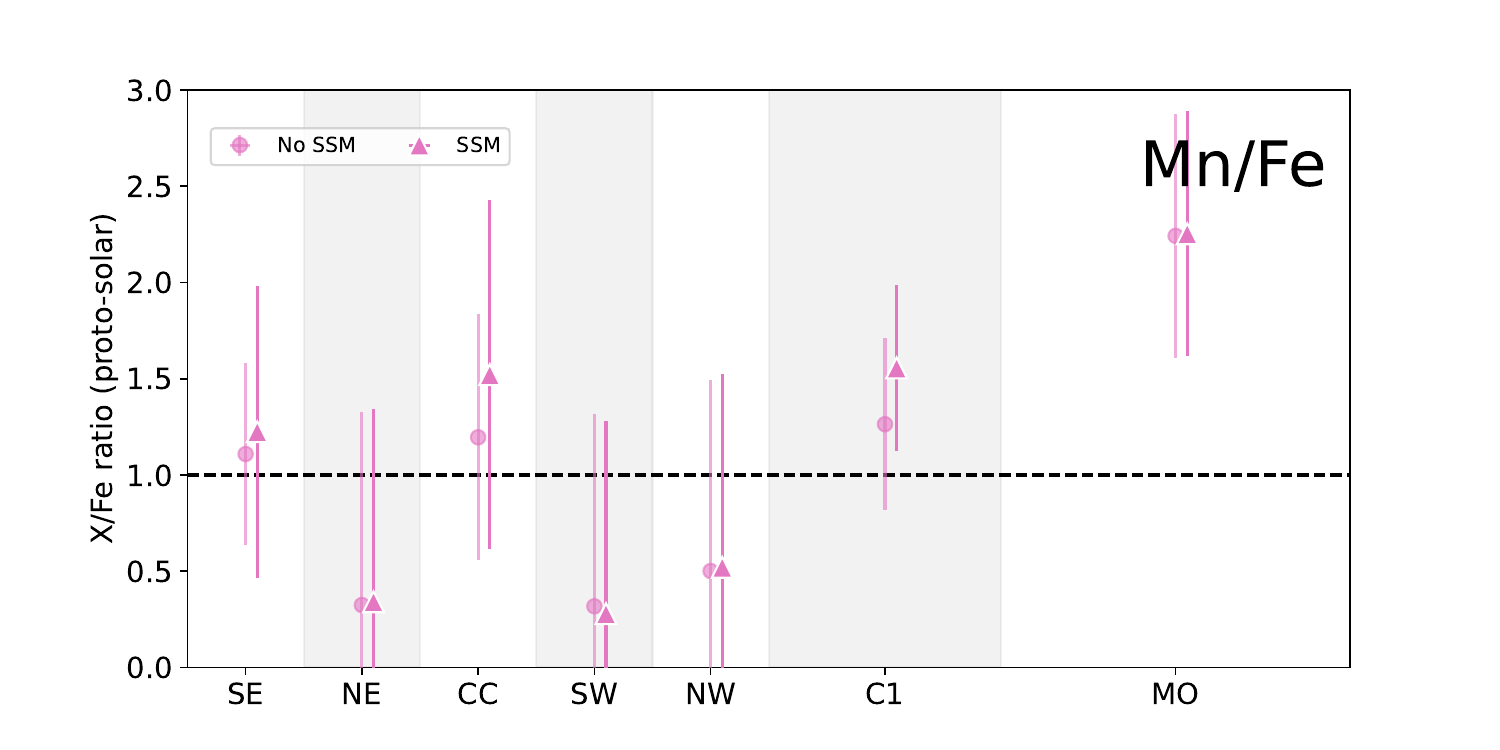} \\
\includegraphics[width=0.32\textwidth, trim={1.5cm 0cm 1.5cm 0cm},clip]{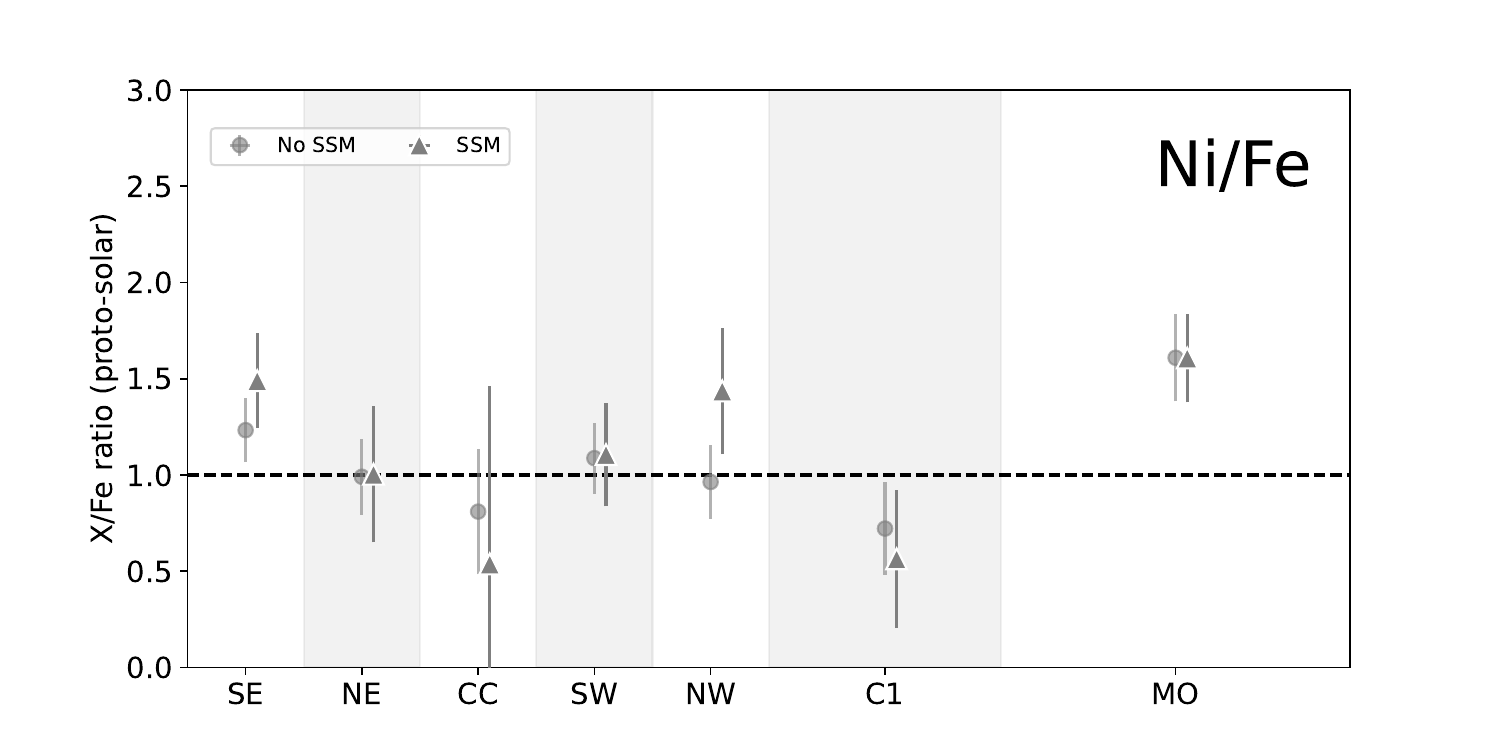} \\
      \caption{X/Fe ratios mapped in all considered regions without and with modelling of the SSM  (Experiments A and B, respectively, of our baseline analysis).}
         \label{fig:ratios}
\end{figure*}

Figure~\ref{fig:ratios} shows the mapping results of each X/Fe ratio accessible with Resolve, with SSM successively ignored and included too. Here again, we find an encouraging consistency between the two approaches, demonstrating that SSM-related biases are marginal. Quite remarkably, all the ratios measured with SSM are $<$2$\sigma$ consistent with solar, with only three exceptions (S/Fe in the CC region, as well as Mn/Fe and Ni/Fe in the MO region). This lack of spatial trend with abundance ratios is further investigated and discussed in Sect.~\ref{sec:discussion:map}.

\begin{table*}[h!]
\caption{Best-fit parameters of spectral analysis (including SSM).}                 
\label{table:abundances}    
\centering                        
\begin{tabular}{l c c c c c c c}      
\hline\hline               
Parameter   &	SE	&	NE	&  CC  & SW   &  NW   &   C1  &   MO \\         
\hline                      

norm ($10^{-2}$~cm$^{-5}$)	&	$6.7 \pm 0.4$	&	$6.7 \pm 0.5$   &	$1.80 \pm 0.24$ &    $8.2 \pm 0.3$ &   $4.5 \pm 0.4$ &   $7.57 \pm 0.17$ &   $3.9 \pm 0.17$ (M1) \\
        	&		&	   &	 &     &    &    &   $1.7 \pm 0.09$ (O1) \\
kT (keV)	&	$3.42 \pm 0.21$	&	$3.54 \pm 0.20$   &	$3.89 \pm 0.13$ &    $3.41 \pm 0.08$ &   $3.30 \pm 0.27$ &   $4.74 \pm 0.08$ &   $6.1 \pm 0.3$ \\
$\sigma_T$ (keV)	&	$2.0 \pm 0.5$	&	$1.7 \pm 0.4$   &	$<0.5$ &    $1.0 \pm 0.3$ &   $1.8 \pm 0.6$ &   $<1.5$ &   $6.2 \pm 0.4$ \\
Fe (solar)	&	$0.71 \pm 0.04$	&	$0.73 \pm 0.05$   &	$ 1.16 \pm 0.19$ &    $0.70 \pm 0.04$ &   $0.77 \pm 0.07$ &   $0.54 \pm 0.03$ &   $0.432 \pm 0.023$ \\
Si/Fe (solar)	&	$1.0 \pm 0.3$	&	$1.0 \pm 0.3$   &	$1.3 \pm 0.5$ &    $0.7 \pm 0.3$ &   $1.1 \pm 0.4$ &   $0.9 \pm 0.4$ &   $<1.2$ \\
S/Fe (solar)	&	$0.65 \pm 0.22$	&	$0.78 \pm 0.24$   &	$1.6 \pm 0.3$ &    $0.85 \pm 0.16$ &   $0.7 \pm 0.3$ &   $1.15 \pm 0.18$ &   $0.88 \pm 0.22$ \\
Ar/Fe (solar)	&	$0.84 \pm 0.25$	&	$0.6 \pm 0.3$   &	$1.3 \pm 0.4$ &    $0.82 \pm 0.20$ &   $<1.2$ &   $1.31 \pm 0.20$ &   $0.9 \pm 0.4$ \\
Ca/Fe (solar)	&	$0.88 \pm 0.18$	&	$0.6 \pm 0.3$   &	$<1.2$ &    $0.96 \pm 0.16$ &   $1.0 \pm 0.3$ &   $1.18 \pm 0.19$ &   $<1.3$ \\
Cr/Fe (solar)	&	$<1.7$	&	$<1.5$   &	$1.9 \pm 0.6$ &    $1.4 \pm 0.4$ &   $<1.7$ &   $0.9 \pm 0.5$ &   $<1.4$ \\
Mn/Fe (solar)	&	$1.2 \pm 0.8$	&	$<1.3$   &	$1.5 \pm 0.9$ &    $<1.3$ &   $<1.5$ &   $1.6 \pm 0.4$ &   $2.3 \pm 0.6$ \\
Ni/Fe (solar)	&	$1.49 \pm 0.25$	&	$1.0 \pm 0.4$   &	$<1.5$ &    $1.1 \pm 0.3$ &   $1.4 \pm 0.3$ &   $0.6 \pm 0.4$ &   $1.61 \pm 0.23$ \\

\hline                                  
\end{tabular}
\end{table*}

\section{Discussion}\label{sec:discussion}

\subsection{The central Fe drop}\label{sec:discussion:drop}

Reported first with \chandra/ACIS by \citet{schmidt2002} and with \XMM/EPIC MOS by \citet{churazov2003}, the central Fe drop in Perseus was then confirmed and further discussed following deeper \chandra observing campaigns of the cluster \citep{sanders2004,sanders2007,fabian2011}. This feature is in fact intriguing as, except a possible depletion of ICM metals into dust \citep{panagoulia2015}, no astrophysical mechanism is known to selectively remove metals in the very core, nor displace them with such azimuthal uniformity.

Despite the uncertainties related to the central AGN emission, our Resolve measurements described in Sect.~\ref{sec:results:systematics} systematically report Fe to be higher than 0.66~solar in the CC region. Combined with the (much better constrained) measurements of Fe $\sim 0.7$~solar in the SE, NE, SW, and NW regions, our results do not corroborate the presence of a strong drop in the core ICM of Perseus.

\begin{figure*}[h!]
\centering
\includegraphics[width=0.42\textwidth, trim={0.2cm 0cm 0.7cm 0cm},clip]{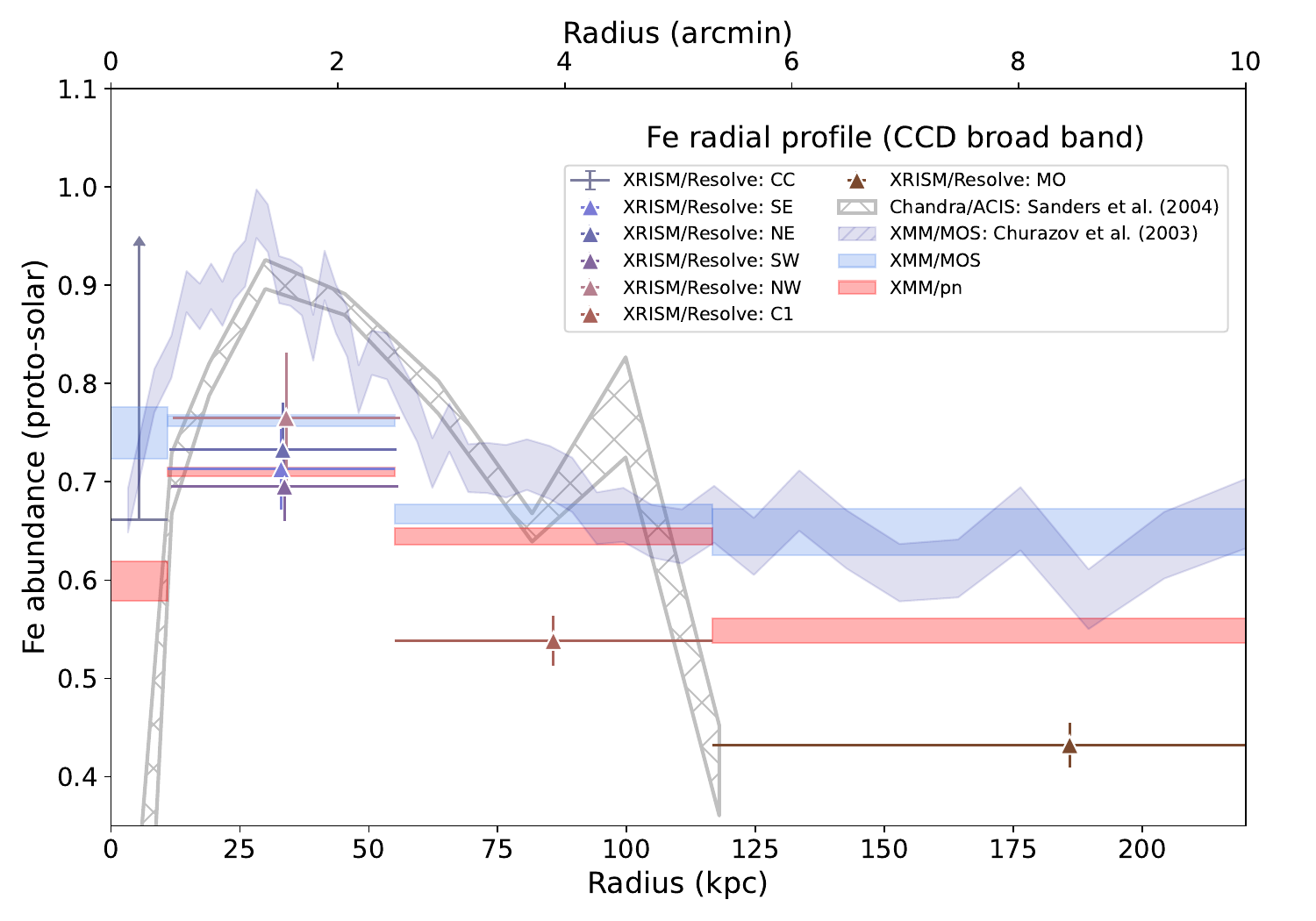}
\includegraphics[width=0.42\textwidth, trim={0.2cm 0cm 0.7cm 0cm},clip]{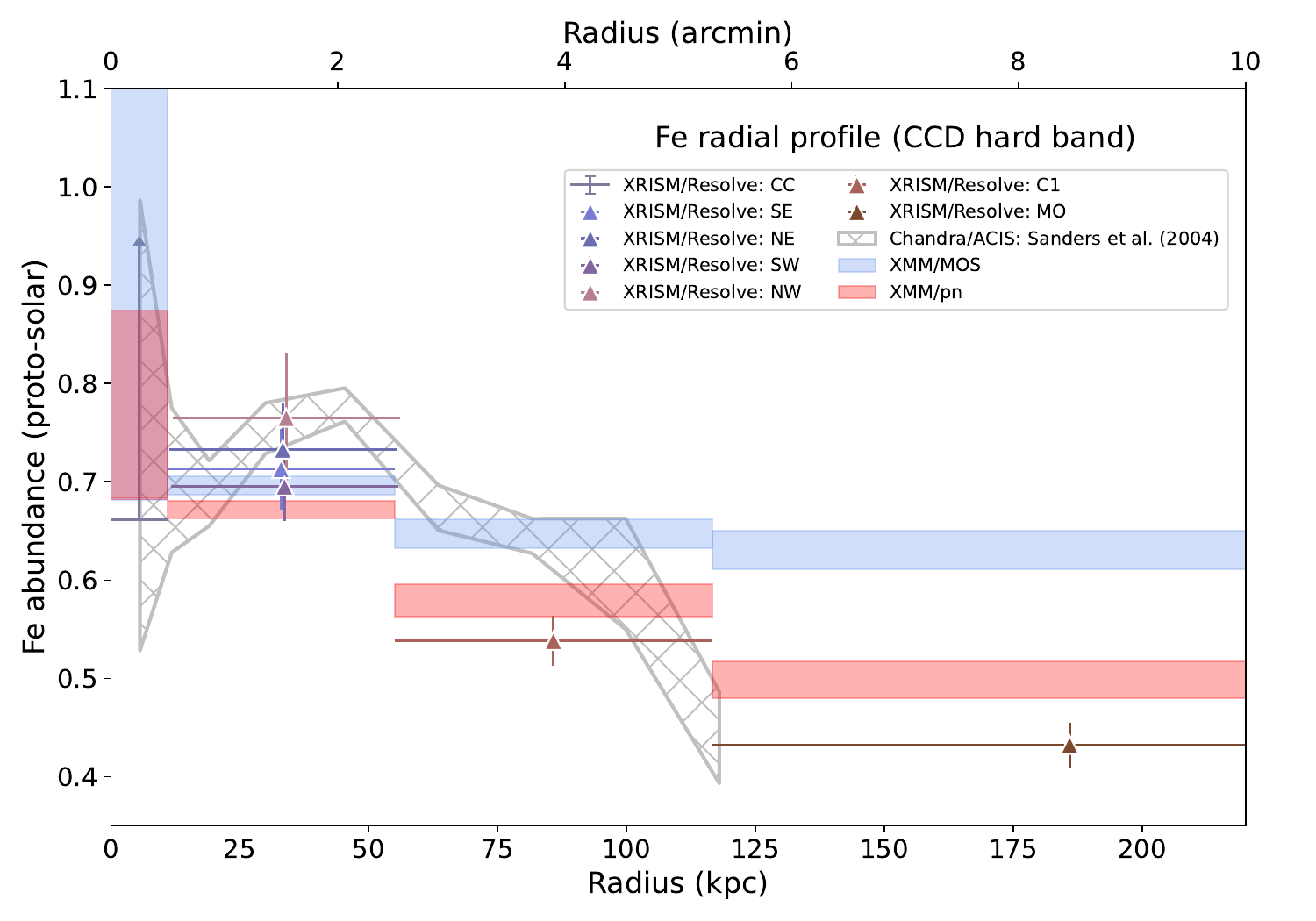}
      \caption{Perseus radial Fe distribution compared between our Resolve analysis (data points) and co-spatial results using \XMM/EPIC MOS and pn instruments (light blue and red boxes). \textit{Left:} Our MOS and pn spectra are fitted over the full energy band (0.45--9~keV). For comparison, we also show the broad-band, azimuthally-averaged Fe profiles from \citet{churazov2003} and \citet{sanders2004}, respectively obtained with MOS and \chandra/ACIS. \textit{Right:} Our MOS and pn spectra are fitted over a harder energy band (1.9--9~keV), selected to better match the Resolve analysis. For comparison, we also show the (azimuthally-averaged) Fe profile from \citet{sanders2004} obtained with ACIS and fitted over 3--8~keV.}
         \label{fig:Fe_comparlit}
\end{figure*}

To clarify this picture in light of moderate-resolution spectroscopy, we show in Fig.~\ref{fig:Fe_comparlit} how our Resolve Fe abundances, re-plotted as a function of their radial distance to the X-ray centroid, compare with similar measurements obtained from \XMM/EPIC MOS (i.e. MOS\,1 + MOS\,2) and pn. We re-extracted a deep EPIC dataset (ObsID:0305780101, $\sim$125~ks net exposure, reduced in the same fashion as in \citealt{mernier2017}) and we analysed its MOS and pn spectra extracted over the same regions as those defined in Sect.~\ref{sec:reduction:region}\footnote{For simplicity, we combined the SW, NE, SW, and NW regions into a single one (i.e. a C0-like pointing excised from the central CC region).}. These spectra were then fitted with the same recipe as our Resolve analysis (gT model, using AtomDB v3.0.9). The EPIC background, largely sub-dominant in all these regions, was estimated from blank-sky observations then subtracted from our raw spectra. As a first exercise, we fit these spectra over a broad energy band, namely 0.45--9~keV\footnote{The 9~keV upper energy limit is considered as an extra precaution to further limit the effect from any imperfect background subtraction on our best-fit parameters.}. To compare these results with the previous literature, we also show in in the left panel of Fig.~\ref{fig:Fe_comparlit} the azimuthally-averaged profiles presented in \citet{churazov2003} and in \citet{sanders2004}, fitted over 0.3--10~keV and 0.6--8~keV, respectively, and rescaled to their \citet{lodders2009} units. All these broad-band CCD profiles are consistent with the presence of a central Fe drop, though with somewhat divergent depth. The comparison also reveals a good agreement between the two MOS analyses. One notable exception is the second innermost bin, where the offset maximum at $\sim$0.9~solar, seen in both \citet{churazov2003} and \citet{sanders2004}, is not recovered in our revisited MOS analysis. We speculate this difference to be (at least partly) explained by the different atomic codes used over the last two decades. In fact, \citet{hitomi2018atom} reported an appreciable decrease of almost 0.1~solar between AtomDB v3.0.2 and v3.0.8 in the same central region of Perseus. 

To allow proper comparison with our Resolve results, we repeat this exercise, now restricting the MOS and pn energy ranges to 1.9--9~keV (Fig.~\ref{fig:Fe_comparlit}, right panel). This second comparison can be completed by an additional ACIS profile from \citet{sanders2004}, fitted also in the hard band (3--8~keV). Interestingly, and unlike the broad-band case above, we now find a reasonable agreement of Fe profiles between Resolve and pn. In contrast, the MOS-pn (or MOS-Resolve) Fe difference is particularly striking in the MO region. Noting the mismatch of mean temperatures between these two instruments ($7.79 \pm 0.15$~keV and $6.26 \pm 0.11$~keV for MOS and pn, respectively), a possible explanation resides in temperature cross-calibration discrepancies found in hot plasmas \citep[see also][]{schellenberger2015}, which must affect the abundances as well. In fact, an overestimated temperature in the MOS fit could artificially increase its Fe abundance parameter to meet the observed line equivalent width at such (biased) temperature. Even more importantly: when excluding the soft band, neither MOS, nor pn, nor ACIS can reproduce the Fe drop, as their abundances in the CC region are highly uncertain too. Already hinted in \citet{sanders2004}, this finding demonstrates that the Resolve band alone is insufficiently suited for characterising metal drops in the ICM. 

An interesting case in this context is that of the Centaurus cluster, famously known for exhibiting a strong Fe drop (e.g. \citealt{sanders2016}; and references therein). Similarly to our results, this drop is not seen in Resolve \citep{fukushima_subm}. That result has been interpreted either as being astrophysical -- for instance the presence of multi-metallicity gas within Centaurus' BCG, NGC\,4696 could shape a drop in the cooler gas phase only \textcolor{blue}{(\citealt{{mernier2026}}; Pl\v{s}ek et al. in prep.)} -- or as a fitting artefact due to a mismodelling of the unresolved Fe-L complex around 1~keV -- for instance due to atomic uncertainties \citep{fukushima2022}. Although the comparison between the two systems is not trivial (the AGN luminosity of NGC\,4696 is known to be negligible), the similar absence (presence) of an Fe drop in the hard Resolve (broad CCD) energy band is striking.

This being said, we cannot fully exclude the existence of an Fe drop in the hot ICM phase probed by Resolve. Statistically speaking, the lower limit of $\sim$0.66~solar seen in the CC region remains $<$2$\sigma$ below the Fe abundance averaged over the SE, NE, SW, and NW regions ($0.717 \pm 0.023$ solar). It should also be noted that the marginal difference between this limit and its corresponding pn broad-band value (whose profile clearly shows a drop) is of the order of $\sim$0.04--0.08~solar at most and might thus be accounted for by unknown residual systematics. More generally, while a detailed cross-instrumental comparison is beyond the scope of this work, we remind the reader that the results discussed in this section are obtained with our current knowledge of the (relatively unconstrained) AGN emission and of the calibration of the telescopes (and instrument). Should this knowledge (and SSM mitigation techniques) evolve with the \XRISM mission lifetime, the above interpretation of the central Fe drop may be revised as well.

\subsection{Spatial uniformity of Perseus' chemical composition}\label{sec:discussion:map}

As introduced in Sect.~\ref{sec:intro}, a key question on cluster (and cluster core) enrichment history is that of the spatial (in)homogeneity of the chemical composition of its ICM. While a number of previous results suggests the latter to be homogeneous \citep{simionescu2009,simionescu2015,mernier2017,ezer2017}, spatial variability of $\alpha$/Fe ratios has been reported in the core of M\,87 \citep{million2011} and in galaxy groups \citep{sarkar2022}. Our \XRISM/Resolve results on Perseus allow us to revisit this picture, for the first time at high spectral resolution. As seen in Fig.~\ref{fig:ratios}, our measurements fully agree with a uniform composition of the gas in all considered regions. We have also demonstrated that this apparent uniformity is not an artificial effect of the \XRISM PSF. Moreover, unlike absolute abundances, X/Fe ratios remain relatively robust toward AGN emission modelling uncertainties (Fig.~\ref{fig:experiments}).

\begin{figure}[h!]
\centering
\includegraphics[width=0.49\textwidth, trim={1.5cm 0.5cm 2.0cm 1.0cm},clip]{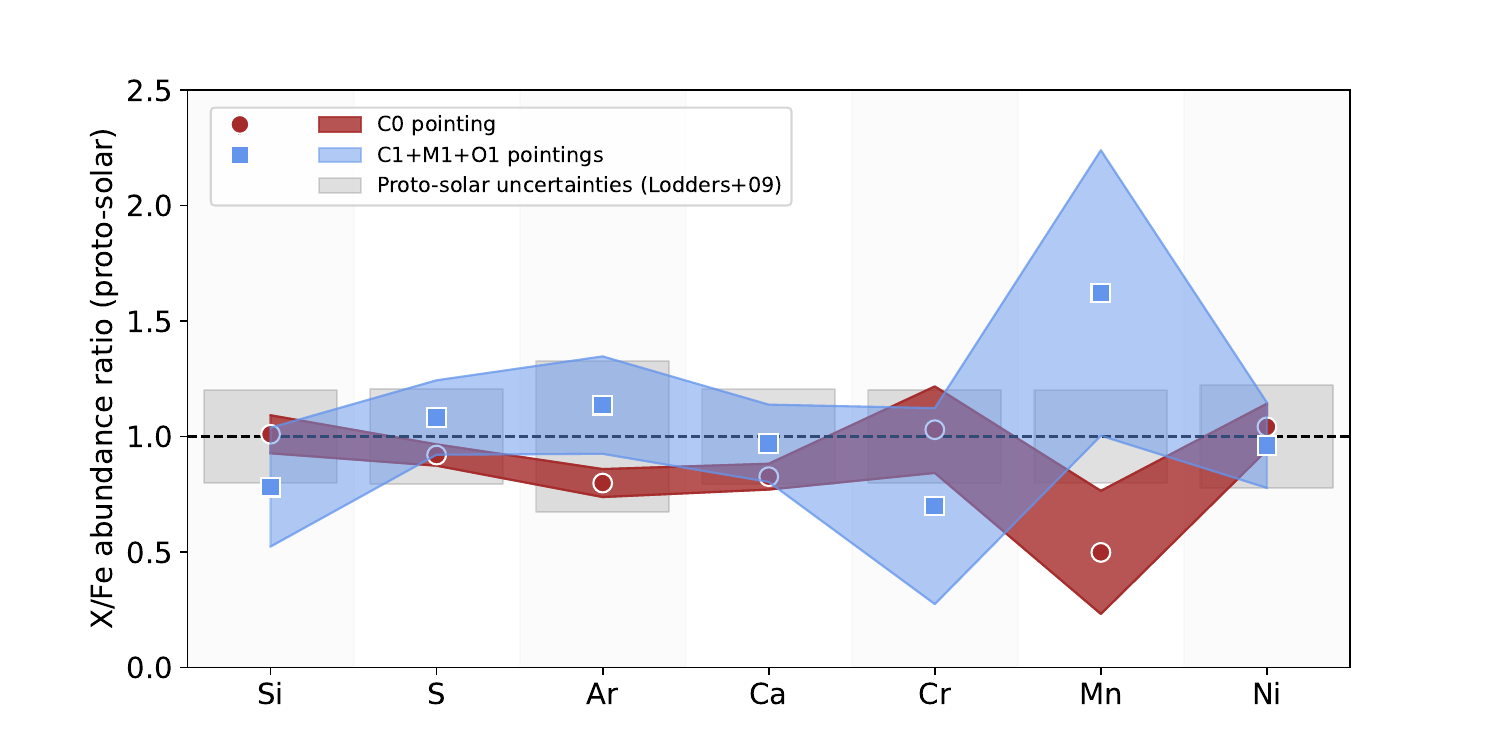} 
      \caption{Abundance pattern averaged over (i) the C0 pointing and (ii) the other (C1+M1+O1) pointings.}
         \label{fig:coarsebin}
\end{figure}

Admittedly, the statistical precision of our measurements in Fig.~\ref{fig:ratios} is limited. The reason is twofold: (i) the choice of relatively small regions in our original mapping strategy and (ii) the error associated to the individual abundances of the X and Fe elements, naturally propagating when we derive its X/Fe ratio. To gain further precision (while keeping SSM into account), we refit our observations over two coarser regions: the entire C0 pointing on one hand and the co-added C1+M1+O1 pointings on the other hand. In the latter, the three spectra (one per pointing) are fitted simultaneously with their norm, $kT$, $\sigma_T$, and Fe abundance parameters allowed to vary individually. In the mean time, we impose each X/Fe ratio to be the same between the three pointings, by coupling their absolute abundances X with Fe, modulo a given factor (the latter representing the actual ratio and being treated as a free parameter). The error calculation on this `direct' X/Fe parameter is now based on the uncertainties of the line flux ratios of X over Fe directly, and do not propagate unnecessarily over those of the continuum. In the rest of this section, this exercise is referred to as the `tied ratios' method. Shown in Fig.~\ref{fig:coarsebin} and Table~\ref{table:final_ratios}, the result confirms the similarity of chemical composition within and beyond $\sim$55~kpc. Except Ar/Fe and Mn/Fe ($<$2$\sigma$), all X/Fe ratios are $<$1$\sigma$ consistent between the two regions.

Since all Fe-peak elements share a common SNIa origin, finding spatially invariant Mn/Fe and Ni/Fe ratios in the ICM is not surprising. On the other hand, $\alpha$ elements (e.g. Si, S, and Ar) are believed to originate at least partly from SNcc contributions. Given that the stellar contribution of NGC\,1275 extends out to $\sim$90~kpc and that the galaxy is mostly red-and-dead, one may expect at first glance a non-negligible excess of late SNIa products (hence lower $\alpha$/Fe ratios) in central regions compared to outer radii. At face value, the dynamics of the central Perseus ICM -- in particular AGN feedback inflating cavities through radio lobes \citep[e.g.][]{gendron2020} and gas sloshing \citep[e.g.][]{walker2018} -- could contribute to erode such initial $\alpha$/Fe gradients. However, such processes should have been equally effective in flattening central Fe peaks, yet unambiguously reported in Perseus and many other relaxed systems as well. Combined with the ability of sloshing cold fronts to preserve (if not enhance) metallicity gradients \citep{ghizzardi2014}, the peaked distribution of absolute abundances makes it unlikely for gas motions solely to explain our flat $\alpha$/Fe profiles. A central excess of SNIa products was in fact the initial interpretation of the central Fe peak in cluster cool cores, since the associated Fe mass of the latter correlates with the mass of the BCG \citep{degrandi2004}. The spatial uniformity of Si/Fe, S/Fe, and Ar/Fe) is thus a valuable result, as it supports a scenario where the majority of SNIa and SNcc end-products were efficiently mixed in the ICM at similar (hence, early) times of cluster formation regardless of its recent central dynamical activity. 

The question of the precise origin of central metal peaks, on the other hand, remains open. The scenario of an ICM pre-enrichment, favoured by our results in Perseus, does not necessarily preclude a profound connection between metal peaks and BGCs, as the latter might have accreted preferentially low-entropy, metal-rich halos throughout their early assembly in clusters cores. Alternatively, one may also be witnessing a coincidental addition of late central SNIa enrichment and outflows emitted directly by the $\alpha$-rich stellar population of NGC\,1275 \citep{ciotti1991,conroy2014}. At first order, this `cosmic conspiracy' scenario would result in a rather similar ICM composition inside and outside cluster cores. Deeper Resolve observations of Perseus and other clusters are necessary to favour one interpretation over the other.

\subsection{Spatially integrated chemical composition}\label{sec:discussion:composition}

\begin{figure*}[h!]
\centering
\includegraphics[width=0.49\textwidth, trim={1.5cm 0cm 1.5cm 0cm},clip]{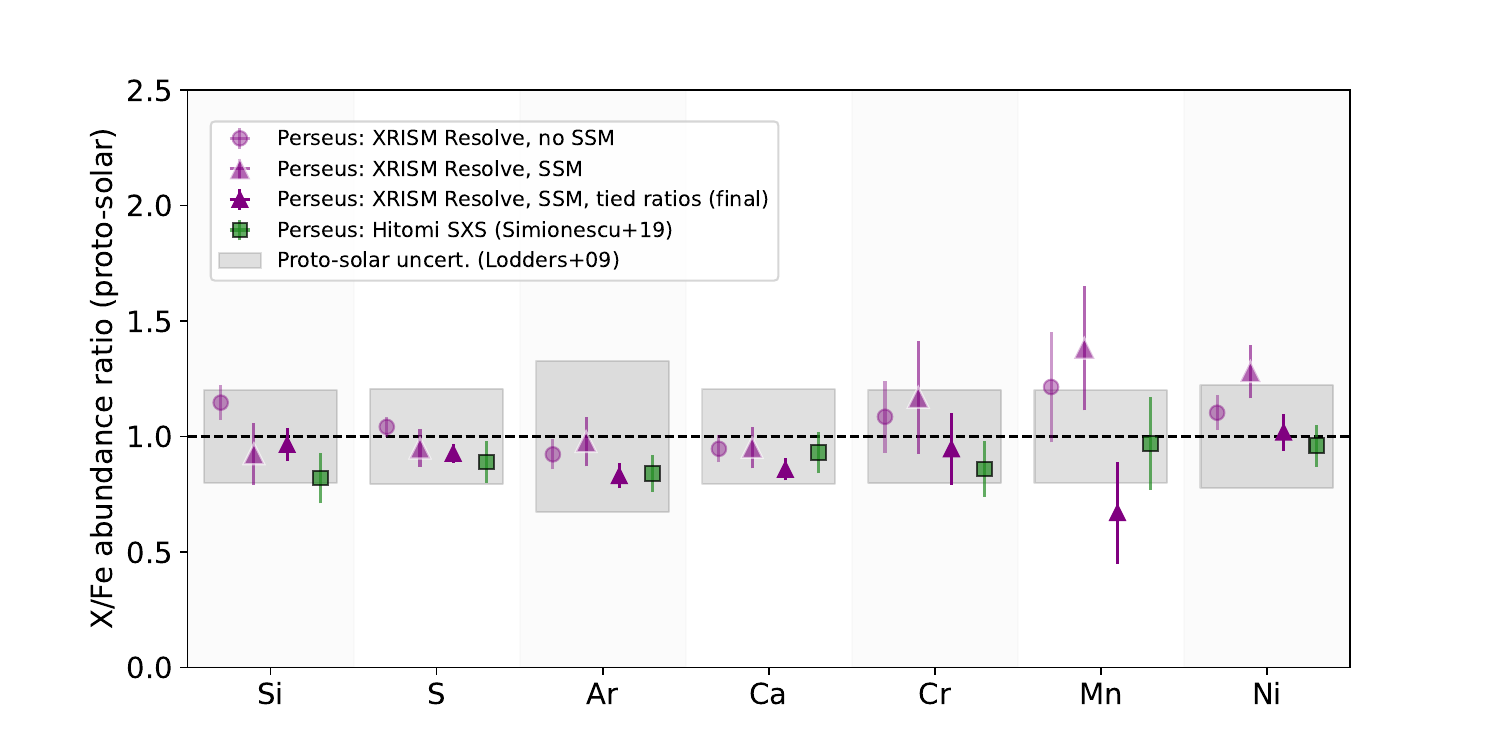} 
\includegraphics[width=0.49\textwidth, trim={1.5cm 0cm 1.5cm 0cm},clip]{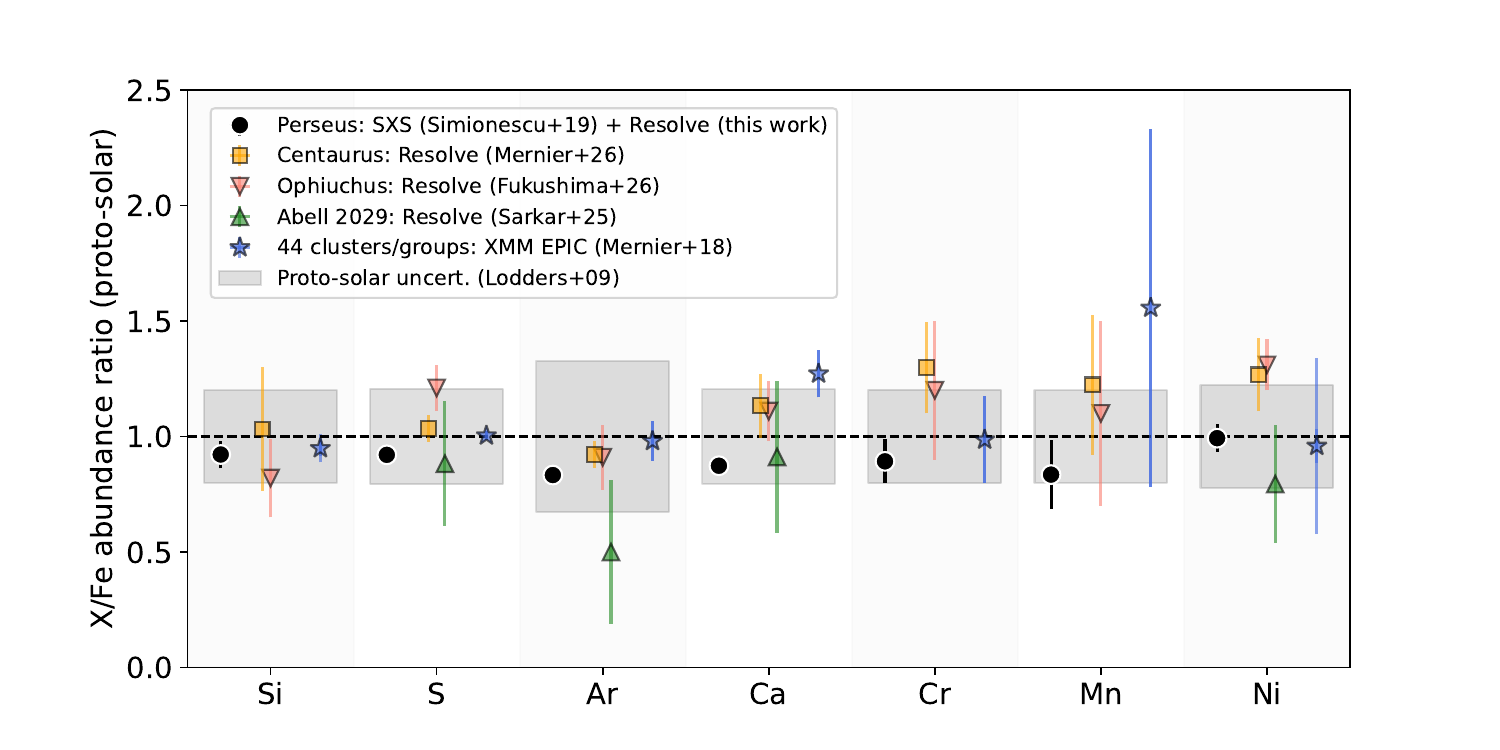} 
      \caption{\textit{Left:} Region-averaged abundance pattern evaluated (i) without SSM and independent abundances in each region, (ii) with SSM and independent abundances in each region, and (iii) with SSM and tied ratios in all regions. These measurements are compared with the \hitomi/SXS ratios obtained by \citet{simionescu2019}. \textit{Right:} Our final (Resolve+SXS) measurements are compared with central Resolve measurements of Centaurus \citep{mernier2026}, Ophiuchus \citep{fukushima2026}, A\,2029 \citep{sarkar2025}, and of EPIC measurements of the CHEERS sample (44 relaxed systems; \citealt{mernier2018}).}
         \label{fig:averaged}
\end{figure*}

Since our results suggest a spatial uniformity of the X/Fe ratios, we can combine each ratio from all regions in order to estimate the chemical composition of the Perseus core as a whole. For consistency, we do so following three separate methods. In the first two, where Gaussian distribution of the errors is assumed, we compute the weighted mean and combined uncertainties of our measurements from Table~\ref{table:abundances} and Fig.~\ref{fig:ratios} (i.e. successively without and with SSM) following the standard formulae:
\begin{equation}
\text{(X/Fe)}_{\text{comb}} =
\frac{\sum_{i=1}^{7} \mathrm{(X/Fe)}_i/\left[\Delta \mathrm{(X/Fe)}_i\right]^2}{\sum_{i=1}^{7} 1/\left[\Delta \mathrm{(X/Fe)}_i\right]^2},
\end{equation}
\begin{equation}
\Delta\text{(X/Fe)}_{\text{comb}} =(\sum_{i=1}^{7} \dfrac{1}{\Delta\text{(X/Fe)}_i^2})^{-\frac{1}{2}},
\end{equation}
where $\mathrm{(X/Fe)}_i$ designates the X/Fe ratio specific to region $i$. We compute these combined numbers for cases with and without SSM successively, as shown in Fig.~\ref{fig:averaged} (light purple triangles and circles, respectively).

Our third method is largely inspired from that of Sect.~\ref{sec:discussion:map}. Starting from the SSM master fit introduced in Sect.~\ref{sec:spectra:SSM}, we modify all our models to tie the X/Fe ratio of every region to one value per element only, following the `tied ratios' method mentioned above. This is equivalent to extracting and fitting one spectrum (with one X/Fe ratio parameter per element X) that covers the four pointings investigated in Sect.~\ref{sec:results:Fe}, yet accounting for the diversity of temperatures, absolute Fe abundances, and dynamical quantities found in the individual regions. As seen in Fig.~\ref{fig:averaged} (dark purple triangles), this experiment provides very precise measurements which, quite remarkably, agree within 1$\sigma$ with all ratios measured with \hitomi/SXS \citep{simionescu2019}. Combining the SXS and Resolve measurements to gain even further precision (Table~\ref{table:final_ratios}), the Perseus abundance pattern can be compared with the ratios averaged over a sample of 44 cool-core systems measured with \XMM/EPIC \citep[the CHEERS sample;][]{mernier2018} and with recent Resolve measurements in the core of the Centaurus cluster \citep{mernier2026}, the Ophiuchus cluster \citep{fukushima2026}, and A\,2029 \citep{sarkar2025}, as shown in the right panel of Fig.~\ref{fig:averaged}. Except a possible tension in Ca/Fe, the agreement between the Perseus ratios and these four datasets is again excellent. Together with other recent \XRISM abundance measurements \citep[e.g. A\,2029; ][]{sarkar2025}, this suggests that, far from being an outlier, the chemical composition of the Perseus core may be representative of that of the central ICM in general (see however the peculiar composition of M\,87 in \citealt{martin2026}).  Although our present results remain confined within the central $\sim$250~kpc ($\sim$0.2~$r_{500}$) of the cluster, ICM abundances are now measured with considerably better accuracy and will allow us in the near future to measure the scatter of central X/Fe ratios in a number of clusters. Meanwhile, the similarity of ratios between the Perseus core (and its remarkably star-forming BCG; e.g. \citealt{mcdonald2018}) on one hand and the average ICM composition from the CHEERS sample on the other hand already suggests that this scatter is limited. At face value, and rejoining the pre-enrichment interpretation discussed in Sect.~\ref{sec:discussion:map}, such a limited scatter may be the result of cluster cool cores having been enriched as early on as the rest of their ICM volume. Nevertheless, in large, closed-box structures such as clusters, one could also consider that systems undergoing different central enrichment histories may eventually converge towards a similar composition over time, since continuous accumulation of metals will naturally tend to average out abundance patterns toward a `central-limit' value (see also discussions in \citealt{simionescu2019}; \citealt{mernier2026}).

\begin{table*}[h!]
\caption{Resolve X/Fe abundance ratios measured over two large regions (Fig.~\ref{fig:coarsebin}) and all regions at once (Fig.~\ref{fig:averaged}), in units of \citet{lodders2009}. }                 
\label{table:final_ratios}    
\centering                        
\begin{tabular}{r | c c c | c}      
\hline\hline               
 Parameter        &	 \multicolumn{3}{c|}{Resolve}     & SXS$^{(a)}$+Resolve 	\\         
    	&	C0    &  C1+M1+O1    &   All         & All 	\\         
\hline                      
Si/Fe (solar)	&	$1.01 \pm 0.08$   &   $0.78 \pm 0.26$   &   $0.96 \pm 0.07$	&	$0.92 \pm 0.05$	\\
S/Fe (solar)	&	$0.92 \pm 0.05$   &   $1.08 \pm 0.16$  &   $0.93 \pm 0.04$	&	$0.92 \pm 0.04$	\\
Ar/Fe (solar)	&	$0.80 \pm 0.06$   &   $1.14 \pm 0.21$   &   $0.83 \pm 0.05$	&	$0.83 \pm 0.04$	\\
Ca/Fe (solar)	&	$0.83 \pm 0.06$   &   $0.97 \pm 0.17$   &   $0.86 \pm 0.05$	&	$0.87 \pm 0.04$	\\
Cr/Fe (solar)	&	$1.03 \pm 0.19$   &   $0.7 \pm 0.4$   &   $0.95 \pm 0.15$	&	$0.89 \pm 0.09$	\\
Mn/Fe (solar)	&	$0.50 \pm 0.27$   &   $1.6 \pm 0.6$   &   $0.67 \pm 0.22$	&	$0.83 \pm 0.15$	\\
Ni/Fe (solar)	&	$1.04 \pm 0.10$   &   $0.96 \pm 0.18$   &   $1.02 \pm 0.08$	&	$0.99 \pm 0.06$	\\

\hline                                  
\end{tabular}
\tablefoot{For each element, the abundance ratio parameters of the combined regions (or pointings) are tied to a unique value.
$^{(a)}$ From \citet{simionescu2019}.
}
\end{table*}

We note that the above considerations remain speculative to some extent, essentially because of the limited number of X/Fe ratios accessible above $E>1.9$~keV. As signature of key $\alpha$-elements produced almost exclusively by SNcc, the O, Ne, and Mg abundances (and their X/Fe counterparts) are of vital importance to obtain a near-complete view of the enrichment channels, processes, and history at play in clusters. For the time being, these lighter abundances can be measured at high resolution only through the \XMM/RGS grating instrument. Already investigated in \citet{simionescu2019} for Perseus, they regrettably suffer from larger uncertainties due to the instrumental broadening of RGS lines in such an extended source. If accomplished eventually, a successful opening of the Resolve gate valve in front of Resolve would provide access to O, Ne, and Mg emission lines at unprecedented resolution and deliver unique constraints on their X/Fe ratios in light of the above discussion.

\subsection{Comparison with SNcc and SNIa yield models}\label{sec:discussion:yields}

The spatially-integrated abundance pattern derived in the previous section is the result of billions of SNcc and SNIa having exploded and enriched the central ICM of Perseus over the last $\gtrsim$10--12~Gyr (Sect.~\ref{sec:intro}). Since clusters behave as closed boxes, their composition bears invaluable information on the bulk of stellar populations having enriched megaparsec-wide structures. Generally speaking, SNcc and SNIa synthesise elemental yields that are highly sensitive to the nature and properties of their progenitors \citep[for a review, see][]{nomoto2013}. Consequently, ICM abundance ratios can in principle help to constrain (i) the time-averaged initial mass function (IMF) and initial metallicity of SNcc progenitors and (ii) the nature (near- vs. sub-Chandrasekhar) and mechanism (deflagration vs. delayed-detonation) of SNIa explosions. 

\begin{figure*}[h!]
\centering
\includegraphics[width=0.33\textwidth, trim={0.2cm 0cm 0.2cm 0cm},clip]{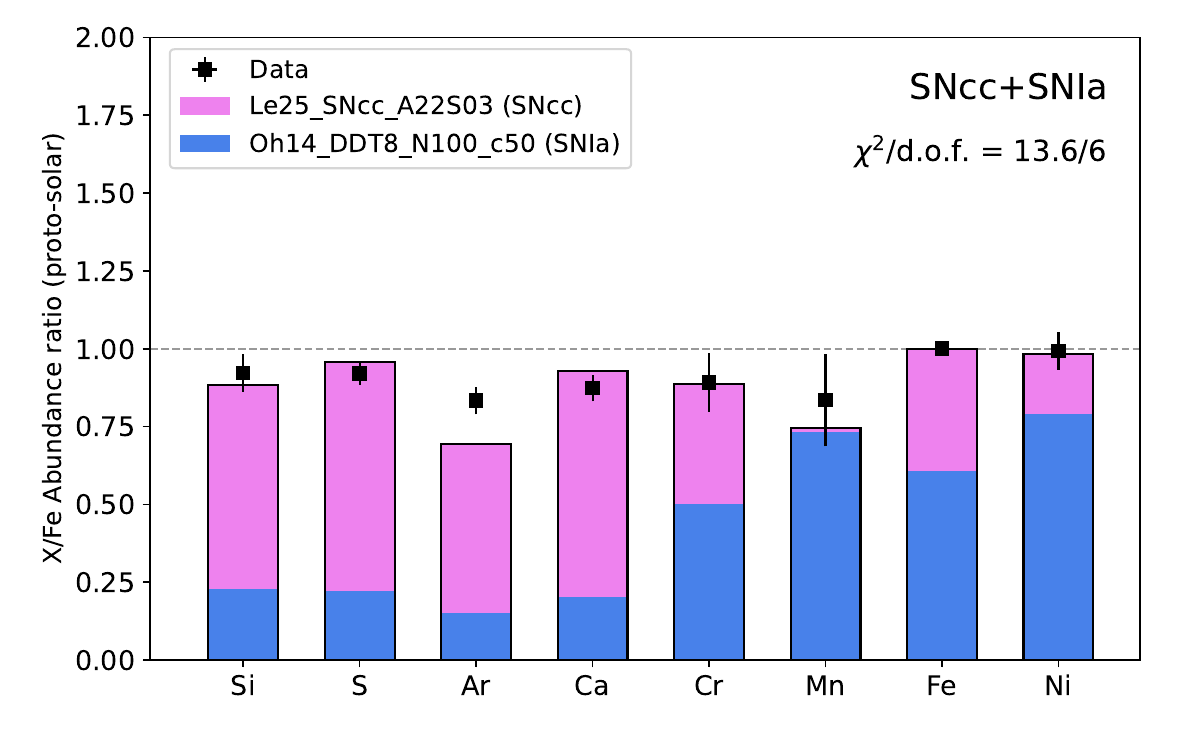} 
\includegraphics[width=0.33\textwidth, trim={0.2cm 0cm 0.2cm 0cm},clip]{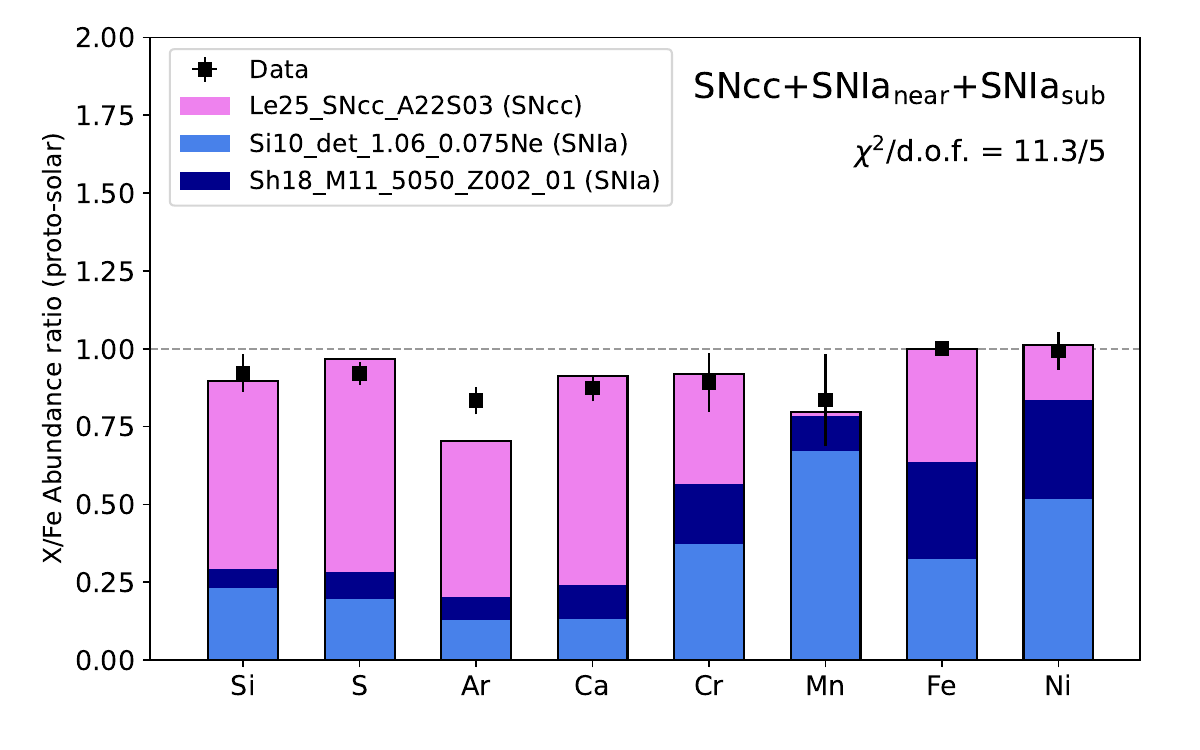} 
\includegraphics[width=0.33\textwidth, trim={0.2cm 0cm 0.2cm 0cm},clip]{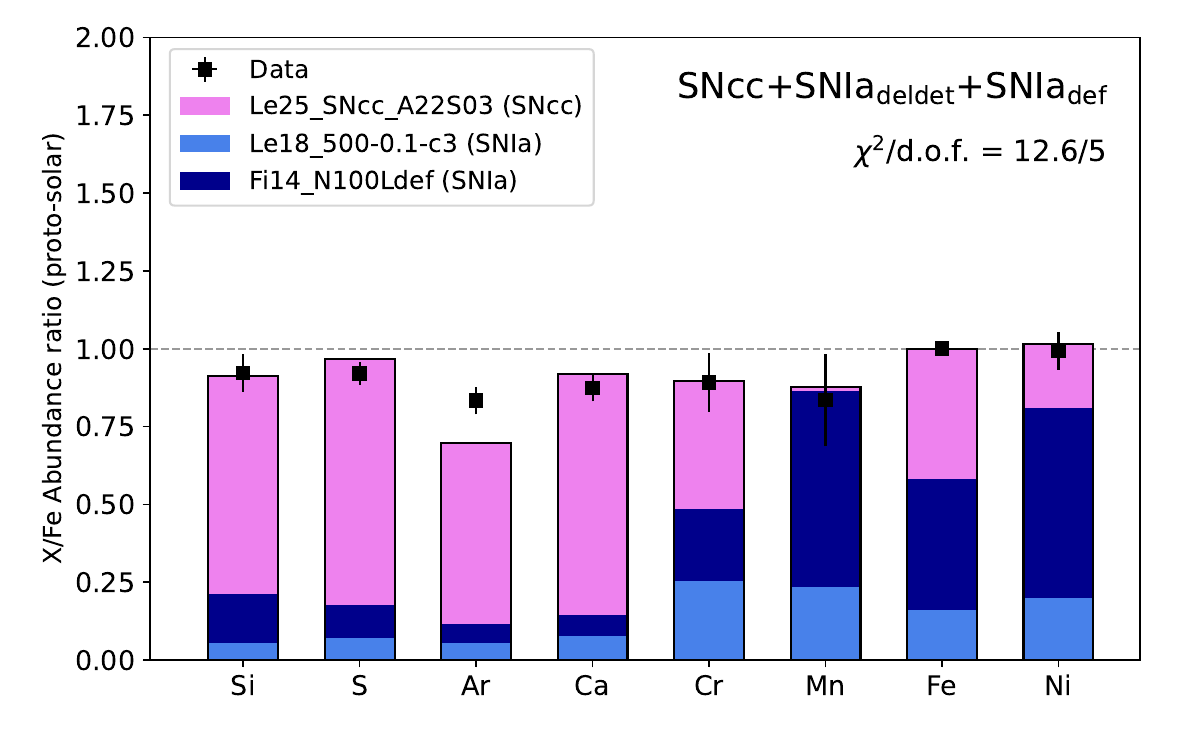}
      \caption{Best-fit combination SNcc and SNIa yield models (colored histograms) on our Perseus final abundance pattern (black data points). SNcc contributions are pink while SNIa contributions are (light and dark) blue. \textit{Left:} Simple SNcc+SNIa model combination. \textit{Middle:} Combination of one SNcc and two SNIa models, assuming co-existence of near- and sub-Chandrasekhar SNIa explosions (SNcc+SNIa$_\mathrm{near}$+SNIa$_\mathrm{sub}$). \textit{Right:} Same as middle panel, now assuming co-existence of deflagration and delayed detonation explosions (in the near-Chandrasekhar scenario).
      }
         \label{fig:SNyields}
\end{figure*}

Inspired by \citet{hitomi2017} and \citet{simionescu2019}, we fit a linear combination of SNcc+SNIa nucleosynthesis yield models to our spatially-integrated (SXS+Resolve) abundance pattern. Using the public code \texttt{abunfit}\footnote{\href{https://github.com/mernier/abunfit}{https://github.com/mernier/abunfit \citep{mernier2016b}}}, we follow the method and strategy described in \citet{mernier2026}, to which we also refer the reader for the complete list of yield models used in this work. While we do not have access to elements entirely produced by SNcc, translating into limited constraints on the yield models of the latter, the precision achieved on the Fe-peak ratios motivates us to investigate the important question of whether Perseus has been enriched by (more than) one SNIa channel. To tackle this question, we perform three successive series of fits: (i) SNcc+SNIa, where only one SNIa model is assumed; (ii) SNcc+SNIa$_\mathrm{near}$+SNIa$_\mathrm{sub}$, where we assume a coexistence of near- and sub-Chandrasekhar SNIa models; and (iii) SNcc+SNIa$_\mathrm{deldet}$+SNIa$_\mathrm{def}$, where we assume a coexistence of delayed-detonation and deflagration SNIa explosions (both in the near-Chandrasekhar scenario). The best-fit combination for each of these three cases is highlighted in Fig.~\ref{fig:SNyields}.

In all three cases, the SNcc model that is being favoured by the fit is SNcc\_A22S03. Proposed recently \citep{leung2025}, this model assumes a Salpeter IMF, solar metallicity of its progenitors, and includes updated calculations of convective process, which have been parametrised to reproduce specifically the Perseus abundance pattern derived by \citet{simionescu2019}. Finding this model to be preferred for our SXS+Resolve dataset is therefore unsurprising and reassuring. Moreover, it demonstrates that our fit does not depend critically on the abundance ratios of lighter elements (O/Fe, Ne/Fe, Mg/Fe) derived at high resolution by RGS only. In fact, the comparison between RGS and Resolve ratios is rendered complicated in our case for a number of reasons -- for instance large instrumental line broadening \citep{simionescu2019}, spatial mismatch of the extraction regions, as well as the possibility of two distinct gas phases with different compositions \citep{mernier2026}; which motivates us to limit our fits to the Resolve ratios. It is interesting to note that the SNcc\_A22S03 model predicts that Si, S, Ar, and Ca are primarily produced by SNcc, making the comparison of SNIa models less sensitive to that of SNcc models. Aside from these encouraging outcome, we also note that the challenge for this SNcc model to reproduce the observed Ar/S ratio persists even after these recent improvements (see discussion in \citealt{simionescu2019}).

Another interesting result is that our three attempts report similar relative fractions of SNIa over the total number of supernovae (SNIa+SNcc): 22\% in the SNcc+SNIa case, 31\% in the SNcc+SNIa$_\mathrm{near}$+SNIa$_\mathrm{sub}$ case, and 35\% in the SNcc+SNIa$_\mathrm{deldet}$+SNIa$_\mathrm{def}$ case, with typical uncertainties of $\sim$4\%. These numbers are comparable with previous ICM estimates \citep[e.g.][]{deplaa2007,mernier2016b,simionescu2019,mernier2026}, with observations of SNIa and SNcc rates \citep{cappellaro1999}, and with theoretical expectations \citep{claeys2014}. However, we caution that they do not necessarily correspond to the actual relative fraction of SNIa having exploded in clusters (nor of SNIa explosion rates), because metals synthesised from these two types may be recycled into stars over different timescales. Instead, they can be seen as the effective contribution of SNIa over the total ICM enrichment \citep[for a detailed discussion, see e.g.][]{deplaa2007,mernier2016b}. Extending this analysis to the two (large) sub-regions of Fig.~\ref{fig:coarsebin}, and assuming the same SNcc+SNIa combination as above, we find that the relative fraction of SNIa is 22\% ($\pm 7\%$) in the C0 region and 17\% ($<$35\%) in the C1+M1+O1 region. The relative contribution of SNIa (SNcc) is thus similar in the two regions, confirming our conclusions of a spatially-invariant ICM composition.

Last but certainly not least, and unlike reported in \citet{hitomi2017} and \citet{simionescu2019}, we do not find statistical preference towards two separate SNIa channels to have enriched the Perseus ICM as the $\chi^2$/d.o.f. values of our fits are very comparable in the three cases. While this different conclusion may be (at least partly) explained by the introduction of the new model SNcc\_A22S03 in our fits (as it contributes to the production of all intermediate elements in a much larger proportion than in previous attempts), we stress that a confirmation of this trend would require a full access to (i) accurate O/Fe, Ne/Fe, and Mg/Fe ratios (see also Sect.~\ref{sec:discussion:composition}), as those would better constrain SNcc models (and in turn, would impact SNIa fits) and (ii) more accurate measurements of Mn/Fe, which is key to discriminate between SNIa models. In particular, the accuracy of Mn/Fe (and other Fe-peak ratios) will be obtained by deeper observations of the Perseus core and/or Resolve measurements averaged over a sample of systems. Such prospects are already being considered by the \XRISM Collaboration and their results will be published in a near future.

\section{Conclusions}\label{sec:conclusion}

In this work, we have presented chemical abundance measurements of the \XRISM/Resolve PV observations of the Perseus cluster. Spread over an arm extending out to $\sim$250~kpc away from the very core, these observations allow to probe the spatial distribution of metals in Perseus for the first time at high ($\sim$5~eV) spectral resolution. Our results can be summarised as follows.

\begin{itemize}
    \item In contrast to previous work based on (broad-band) CCD measurements, we do not see evidence for a central Fe abundance drop in Perseus. Specifically, the Fe abundance in the inner $\sim$10~kpc is systematically measured to be consistent with, or higher than that of its immediate surroundings, regardless of the wide range of priors we have considered in our spectral analysis. Although this apparent lack of Fe drop has also been observed with Resolve in the Centaurus cluster and might reveal a genuine metal peak in the hotter phase of the ICM, the uncertain contribution of the central AGN and the imperfect knowledge of the effective area calibration at sub-array scales prevent us from firmly ruling out the possibility of a (shallow) metal drop in Perseus.
    \item Considerably less sensitive to the central AGN contribution, the Si/Fe, S/Fe, Ar/Fe, Ca/Fe, Cr/Fe, Mn/Fe, and Ni/Fe ratios are measured to be consistent with their solar value in all investigated regions. In line with a number of previous CCD results, this spatial invariance can be explained in the context of a pre-enrichment of the central ICM, with negligible contribution from recent ejection of metals by Perseus' BCG, NGC\,1275. Favouring this scenario over that of a cosmic conspiracy, in which late SNIa products and $\alpha$-rich stellar outflows would mimic the solar, pre-enriched composition of the outer ICM, would require deeper observations and access of O, Ne, and Mg lines below 2~keV.
    \item Integrating the ratios over all regions probed by Resolve allows to measure the (near-solar) composition of the Perseus core with excellent statistics. Combined with (fully consistent) measurements from \hitomi/SXS observations, we compare the Perseus abundance pattern with combinations of SNcc and SNIa nucleosynthesis yield models from the literature, with the aim to provide astrophysical constraints on the latter. Unlike what has been suggested in previous work, we do not find evidence for a need of more than one SNIa channel to explain the Perseus enrichment.
\end{itemize}

While we believe that these conclusions are of significant importance for the community, we stress again the importance of \XRISM observations to accumulate, calibration knowledge to mature, and dedicated analysis techniques to develop in order to refine the results (and thus the interpretations) presented in this paper.

\begin{acknowledgements}
      This paper is dedicated to the memory of our colleague Katja Pottschmidt who passed away on June 17, 2025. Besides her invaluable expertise on the \XRISM and \nustar missions and on the astrophysics of compact objects, her kindness and generosity will be dearly missed among the X-ray astronomical community. The results presented above are made possible by over three decades of work by the team of scientists and engineers who created a microcalorimeter array for X-rays and overcame enormous setbacks. We gratefully acknowledge the entire \XRISM team's effort to build, launch, calibrate, and operate this observatory. We thank the referee for useful suggestions which helped to improve the quality of this manuscript. We thank Eugene Churazov and Jeremy Sanders for kindly providing observational data from previous work.
Part of this work was supported by the U.S.\ Department of Energy by Lawrence Livermore National Laboratory under Contract DE-AC52-07NA27344, and by 
NASA under contracts 80GSFC21M0002 and 80GSFC24M0006 and grants 80NSSC20K0733, 80NSSC18K0978, 80NSSC20K0883, 80NSSC20K0737, 80NSSC24K0678, 80NSSC18K1684, 80NSSC23K0650, and 80NNSC22K1922.
Support was provided by JSPS KAKENHI grant numbers JP23H00121, JP22H00158, JP22H01268, JP22K03624, JP23H04899, JP21K13963, JP24K00638, JP24K17105, JP21K13958, JP21H01095, JP23K20850, JP24H00253, JP21K03615, JP24K00677, JP20K14491, JP23H00151, JP19K21884, JP20H01947, JP20KK0071, JP23K20239, JP24K00672, JP24K17104, JP24K17093, JP20K04009, JP21H04493, JP20H01946, JP23K13154, JP19K14762, JP20H05857, JP25K23398, and JP23K03459, the JSPS Core-to-Core Program, JPJSCCA20220002, and the Strategic Research Center of Saitama University.
FM acknowledges financial support from the Centre national d’études spatiales (CNES), France (ROR: https://ror.org/04h1h0y33) within the framework of the \textit{XRISM} mission. 
EB is supported by The Israel Science Foundation (grant No. 2617/25). 
LC acknowledges support from NSF award 2205918. 
CD acknowledges support from STFC through grant ST/T000244/1. 
LG acknowledges support from Canadian Space Agency grant 18XARMSTMA.
NO acknowledges partial support by the Organization for the Promotion of Gender Equality at Nara Women's University. 
MS acknowledges support by the RIKEN Pioneering Project Evolution of Matter in the Universe (r-EMU) and Rikkyo University Special Fund for Research (Rikkyo SFR). 
AT acnowledges support from the Kagoshima University postdoctoral research program (KU-DREAM). 
SU acknowledges support by Program for Forming Japan's Peak Research Universities (J-PEAKS).
SY acknowledges support by the RIKEN SPDR Program. 
IZ acknowledges partial support from the Alfred P.\ Sloan Foundation through the Sloan Research Fellowship.
CZ acknowledges the support of the Czech Science Foundation (GACR) Junior Star grant no. GM24-10599M.
\end{acknowledgements}

\bibliographystyle{aa} 
\bibliography{aa59603-YY}

\begin{appendix}

\section{Spatial-spectral mixing region coefficients}\label{sec:app:SSMcoef}

In this Appendix section, we provide a list of fractions $f_{J\rightarrow i}$ of photons leaking from sky region $J$ into detector region $i$, as described in Sect.~\ref{sec:spectra:SSM}. These coefficients are shown in Table \ref{table:SSMcoef}. We stress that sky regions $J$ often differ in extent from their corresponding detector regions $j$ (to account for external SSM; see also Fig.~\ref{fig:map}), which explains why the diagonal coefficients $f_{J\rightarrow j}$ are considerably smaller than unity. Although MO is considered as a single region in our analysis, it consists of the separate M1 and O1 Resolve pointings. Therefore, we treat these two regions individually in this exercise.

\begin{table}[h!]
\caption{$f_{J\rightarrow i}$ fractions of leaking photons.}   
\label{table:SSMcoef}    
\centering                        
\begin{tabular}{l | c c c c c c c c}      
\hline\hline               
Detector $\rightarrow$& se & ne & cc & sw & nw & c1 & m1 & o1 \\         
Sky  $\downarrow$    &  &  &  &  &  &  &  \\         
\hline                      

SE & 0.35 & 0.05 & 0.05 & 0.03 & 0* & 0* & 0* & 0* \\
NE & 0.08 & 0.32 & 0.07 & 0* & 0.08 & 0* & 0* & 0* \\
CC & 0.14 & 0.14 & 0.38 & 0.08 & 0.10 & 0* & 0* & 0* \\
SW & 0.05 & 0* & 0.05 & 0.30 & 0.05 & 0.05 & 0* & 0* \\
NW & 0* & 0.06 & 0.06 & 0.05 & 0.36 & 0.18 & 0* & 0* \\
C1 & 0* & 0* & 0* & 0.01 & 0.09 & 0.51 & 0.04  & 0* \\
M1 & 0* & 0* & 0* & 0* & 0* & 0.09 & 0.37  & $-$ \\
O1 & 0* & 0* & 0* & 0* & 0* & 0* & $-$ & 0.38 \\

\hline                                  
\end{tabular}
\tablefoot{Coefficients of non-adjacent regions are assumed to be negligible (0*). The M1 and O1 pointings are part of the same region MO, therefore no coefficient is applicable ($-$).
}
\end{table}

\end{appendix}
\end{document}